%% file: mnras_canelo_revised3.tex
\documentclass[a4paper,fleqn,usenatbib]{mnras}

\usepackage[T1]{fontenc}
\usepackage{ae,aecompl}

\usepackage{graphicx}	
\usepackage{caption,xtab,booktabs} 
\usepackage{subfigure} 

\title[The 6.2~$\mu$m profile in starburst-dominated galaxies]{Variations in the 6.2~$\mu$m emission profile in starburst-dominated galaxies: a signature of polycyclic aromatic nitrogen heterocycles (PANHs)?}

\author[C. M. Canelo et al.]{
C. M. Canelo,$^{1}$\thanks{E-mail: camcanelo@gmail.com}
A. C. S. Fria\c{c}a,$^{1}$
D. A. Sales,$^{2}$
M. G. Pastoriza$^{3}$
and D. Ruschel-Dutra$^{4}$
\\
$^{1}$Departamento de Astronomia, Instituto de Astronomia, Geof\'isica e Ci\^encias Atmosf\'ericas, Universidade de S\~ao Paulo, S\~ao Paulo, Brazil\\
$^{2}$Instituto de Matem\'atica, Estat\'istica e F\'isica, Universidade Federal do Rio Grande, Rio Grande do Sul, Brazil \\
$^{3}$Instituto de F\'isica, Universidade Federal do Rio Grande do Sul, Rio Grande do Sul, Brazil \\
$^{4}$Centro de F\'isica e Matem\'atica, Universidade Federal de Santa Catarina,  Santa Catarina, Brazil
}

\date{Accepted XXX. Received YYY; in original form ZZZ}

\pubyear{2017}

\begin{document}
\label{firstpage}
\pagerange{\pageref{firstpage}--\pageref{lastpage}}
\maketitle

\begin{abstract}
Analyses of the polycyclic aromatic hydrocarbon (PAH) feature profiles, especially the 6.2~$\mu$m feature, could indicate the presence of nitrogen incorporated in their aromatic rings. In this work, 155 predominantly starburst-dominated galaxies (including HII regions and Seyferts, for example), extracted from the Spitzer/IRS ATLAS project \citep{caballero}, have their 6.2~$\mu$m  profiles fitted allowing their separation into the Peeters' A, B and C classes \citep{Peeters02}. 67\% of these galaxies were classified as class A, 31\% were as class B and 2\% as class C.  Currently class A sources, corresponding to a central wavelength near 6.22~$\mu$m, seem only to be explained by polycyclic aromatic nitrogen heterocycles \citep[PANH,][]{Hud05}, whereas class B may represent a mix between PAHs and PANHs emissions or different PANH structures or ionization states. Therefore, these spectra suggest a significant presence of PANHs in the interstellar medium (ISM) of these galaxies that could be related to their starburst-dominated emission. These results also suggest that PANHs constitute another reservoir of nitrogen in the Universe, in addition to the nitrogen in the gas phase and ices of the ISM.
\end{abstract}




\section{Introduction}
\label{intro}

A considerable fraction of the carbon in the interstellar medium (ISM), $20\%$ or more, is in the form of Polycyclic Aromatic Hydrocarbons (PAHs) \citep{job92}. In addition, the mid-infrared (MIR) emission from many objects is dominated by bands of a molecular class, that includes PAHs, sometimes referred to the Aromatic Infra-red Bands (AIBs) \citep{job92}. All other classes of organics and inorganics represent only a tiny fraction of the emitting material that contributes to the AIBs \citep{allamandola99}. Up to 50$\%$ of the luminosity emitted in the MIR can be due to PAHs, with the most prominent bands emitting at 3.3, 6.2, 7.7, 8.6, 11.3 and 12.7~$\mu$m \citep{Li04}. Their high luminosity allow them to be observed in high redshift objects, where they may dominate the IR spectrum range  \citep{Papovich06,  Teplitz07}. Recently, the largest redshift in which PAH bands were detected was for the Cosmic Eye Galaxy with z~$=3.074$, a Lyman Break Galaxy with strong gravitational lens \citep{Siana}. Years later, the 6.2~$\mu$m band was observed in the submillimeter galaxy GN20, with a redshift of z~$=4.055$ \citep{riechers14}.

Because of their stable molecular structure, PAHs are the dominant molecular organic material in space \citep{Eh06} and, together with other aromatic macromolecules, they are the most abundant class of molecular species that must have been transported to the planets by comets, meteorites and interplanetary dust deposition \citep{Eh02}. Produced in other parts of the Solar System or Galaxy, they have been delivered almost intact to planets such as Earth and Mars. They could also have undergone a stage of production of nitrogen heterocyclic molecules which, along with PAHs, are of a great astrobiological interest. In addition, in the PAH World model for the origins of life, they played a key role in the stages preceding the RNA World \citep{Eh06}, not only on Earth but in other astrophysical environments as well.

When a PAH incorporates nitrogen in place of a carbon atom, it is called a  polycyclic aromatic nitrogen heterocycle (PANH). It has been suggested that a significant fraction of the nitrogen  in the ISM is depleted into PANHs \citep{Hud05}. \citet{Peeters02} considered the 6 -- 9~$\mu$m spectral range of several astrophysical objects and studied the presence of profile variations among the PAH bands. They found that their sample could be separated into three different classes -- A, B and C --  depending on the peak positions of the bands. Later, \citet{died04} extended the approach for the 3.3~$\mu$m and 11.2~$\mu$m bands and revealed a correlation between the classification of the PAH bands and their profiles. 

In general, an A classification in the 6 -- 9~$\mu$m region also implies an A$_{3.3}$ and A$_{11.2}$, but B$_{3.3}$ and B$_{11.2}$ do not necessarily correlate with each other or with  B$_{6-9}$ and C$_{6-9}$. Normally, the profile A peaks at 6.2~$\mu$m while profile B and C peak at longer wavelengths. The classes A and B differ largely in the relative strength of subcomponents at 7.6 and $7.8 \mu m$, which seem to have shifted to $8.2 \mu m$ for class C \citep{Tielens08}. Also, the classes are linked to the type of source. Class A sources are associated with interstellar material illuminated by a star, including HII regions, reflection nebulae, and the general ISM of the Milky Way and other galaxies. Class B objects are associated with circumstellar material and include planetary nebula, a variety of post-AGB objects and Herbig AeBe stars. Class C sources are limited to a few extreme carbon-rich post-AGB objects.

Thus, analyses of PAH feature profiles, especially the 6.2~$\mu$m band, could indicate the presence of nitrogen incorporated into the rings. The class A 6.2~$\mu$m band corresponds to a central wavelength at 6.22~$\mu$m and has only been well reproduced by carbon replaced by nitrogen into the aromatic rings \citep{Hud05}. These PAH features are prominent in star-forming systems, reduced and modified in high-intensity starbursts and, eventually, disappear in active galactic nuclei (AGN) systems \citep{yan07}. The spectral continuum shape of starburst spectra is dominated by strong emission features from PAHs \citep{Genzel00} and the 5 -- 8~$\mu$m spectral range of starburst galaxies is not only extremely rich in atomic and molecular emission and absorption features but is also dominated by emission from the 6.2~$\mu$m PAH feature and the blue wing of the 7.7~$\mu$m PAH complex \citep{Brandl06}. In fact, the emission and absorption features of dust grains predominate the MIR spectra of starburst galaxies and most ULIRGs \citep[Ultra-Luminous Infrared Galaxies, ][]{yan05}.

Identification of the feature classes can show if PANHs may be present in the spectrum of the sources and account for this PAH band of the MIR emission. With this in mind, we here analyzed and classified the 6.2~$\mu$m feature of 206 galaxies observed with Spitzer according to the Peeters' classes, searching for the PANH contribution to this PAH band. This paper is structured as follows: Section \ref{data} explains the selection of our sample and Section \ref{analysis} describes the data analysis performed in the spectroscopic data. Results are discussed in Section \ref{results} and Section \ref{conclusion} presents the summary and conclusion.


\section{Data selection}
\label{data}

The Spitzer/IRS ATLAS project \citep{caballero} possess around 750 reduced spectra of several types of extragalactic objects, such as Seyfert, radiogalaxies and submillimeter galaxies. They were observed in low and high resolution by the Infrared Spectrograph \citep[IRS, ][]{Houck04} of the Spitzer Space Telescope \citep{Werner04} and their reduced spectra were extracted from Postscript figures uploaded to the arxiv.org preprint service by their authors. The ATLAS project also offers the best redshift values of the sources found in the literature and checked with NED (NASA Extragalactic Database). The data are not at the rest wavelength and, for this correction, we used values published by the authors (see Table~\ref{tab:sources}). However, in the case of high redshifts, the values could have been obtained only through PAH bands themselves or through the IRS spectrum, with no other accurate measure. According to \citep{Weedman06b}, the uncertainties for spectrostopic redshifts greater than 1 from IRS spectrum are typically $\pm$~0.2. These sources are indicate in the table. This situation may interfere with the rest wavelengths of the spectra and compromise the original peak position of the features. For this reason, these cases must take into account a possible extra shift in the 6.2~$\mu$m central wavelength for the analysis. 
	
The spectra may have an inferior quality if compared to the fully and properly reduced spectra because, in some cases, the flux uncertainties are missing or the spectra are smoothed, for instance. The accuracy with which the original wavelength and flux values are recovered is limited by the resolution of the Postscript figure, once they were transformed from the Postscript coordinates (sets of points representing the spectra in the figures) \citep{caballero}. Nevertheless, according to the authors, the resulting introduced uncertainty to the wavelength calibration is an order of magnitude smaller than the spectral resolution (R~$\sim$~100) in the low-resolution module of IRS. Therefore, its impact is negligible and statistical analysis of the sample is little, if at all, affected \citep{caballero}.
 	 
For a better understanding of the PANH distribution in the Universe, starburst galaxies are the best targets since they carry different burst of young stellar population and, consequently, present strong PAH emission in the MIR spectral wavelengths. We therefore selected objects from the ATLAS MIR starburst-dominated galaxies (MIR\_SB sample), which is composed of 257 sources previously classified as starburst-dominated by \citet{caballero}. The limit set between AGN- and starburst-dominated sources was based on the fraction of a PDR (photo-dissociation region) component at $r_{PDR}=0.15$, corresponding to equivalent widths (EW) of EW$_{6.2}=0.2 \mu m$ or EW$_{11.3}=0.2 \mu m$ as an alternative boundary.

\begin{figure}
\centering
\includegraphics[scale=0.45]{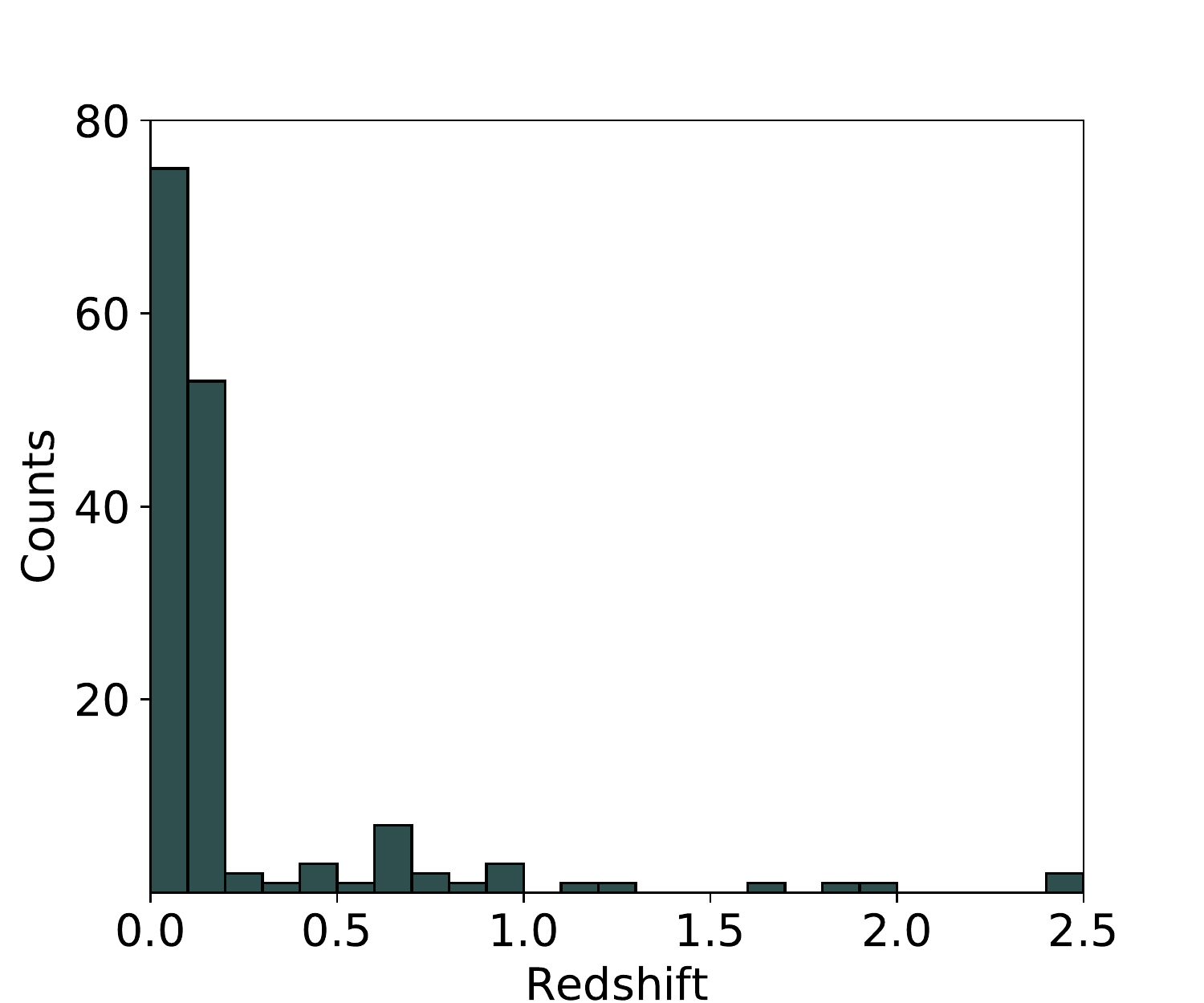}
\caption{Histogram of the 155 redshifts of our galaxy's sample.}
\label{fig:hist1}
\end{figure}

From their sub-sample, 219 objects have a wavelength coverage that includes the 6.2~$\mu$m PAH band.  However, some galaxies do not have enough data points for a trustworthy fit or the profiles were peculiar and were not used in this work. We also included 2 other ULIRGs observed by \citet{yan07} due to their strong PAH features. Table~\ref{tab:sources} presents information of our sample that is composed by 155 galaxies. The distribution of the redshifts can be seen in Fig. \ref{fig:hist1}.

\section{Data Analysis}
\label{analysis}

Before the 6.2~$\mu$m feature profile was fitted, the spectral contributions of the silicate absorption and line emissions were subtracted from the spectra using PAHFIT (Smith et al., 2007). This IDL\footnote{Interactive Data Language, available at http://ittvis.com/idl} script was created to decompose low resolution IRS spectra into dust features, stellar and thermal dust continuum, silicate absorption and ionic and molecular line emission. Although PAHFIT also recovers PAH features, the central wavelengths of the bands are fixed in the code. This lack of flexibility  prevents the account of the peak position variations and the fits of PAH bands obtained with this tool were disregarded in this work.

The continuum of the galaxies was fitted with a spline with anchor points at roughly 5.0, 5.4, 5.5, 5.8, 6.6, 7.0, 8.2, 9.0, 9.3, 9.9, 10.2, 10.5, 10.7, 11.7, 12.1, 13.1, 13.9, 14.7 and 15.0~$\mu$m according to the method utilized in \citet{Peeters17}. The inclusion of each point depended on the presence of the PAH plateaus (at 5--10~$\mu$m and 10--15~$\mu$m) and molecular bands (at 10.68~$\mu$m, for example). The spline decomposition was chosen because it allows to isolate the 6.2~$\mu$m band.

\begin{figure}
\centering
\resizebox{\hsize}{!}{\includegraphics{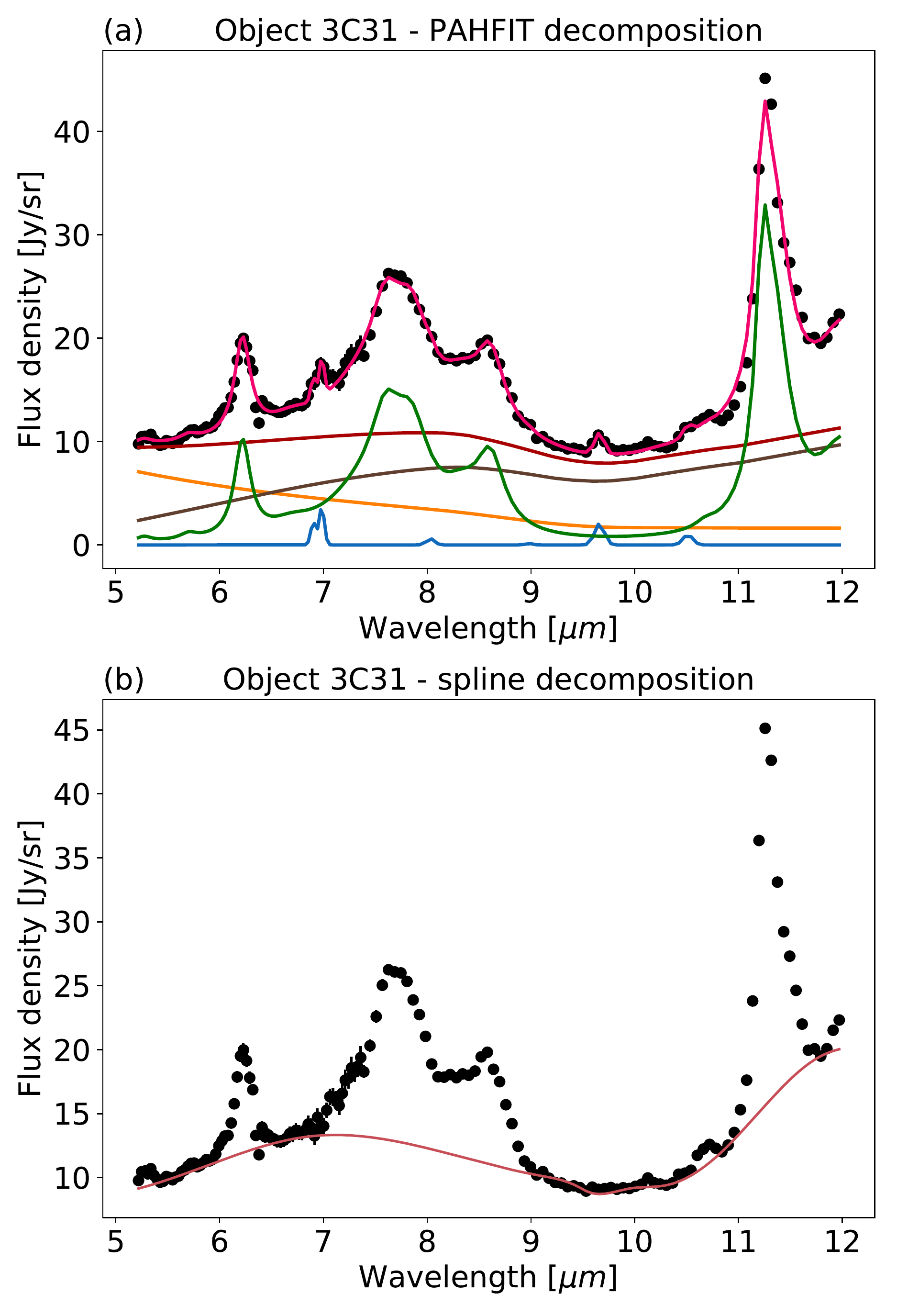}}
\caption{Comparison between the spectral decomposition with PAHFIT and with spline continuum for an object with the PAH plateaus.  (a) Result of the PAHFIT decomposition of the 3C31 spectrum. The data points are represented by the dots with the vertical error-bars as uncertainties.  The pink line corresponds to the best fit model, the green line is the dust (AIB, PAH) contribution and blue one is the ionic and molecular lines contribution. The red line represents the total continuum contribution, and the brown and orange lines correspond to its individual thermal and stellar components, respectively. (b) The red line represents the spline decomposition of the continuum. The data points are represented by the dots with the vertical error-bars as uncertainties.}
\label{fig:pahfit1}
\end{figure}

\begin{figure}
\centering
\resizebox{\hsize}{!}{\includegraphics{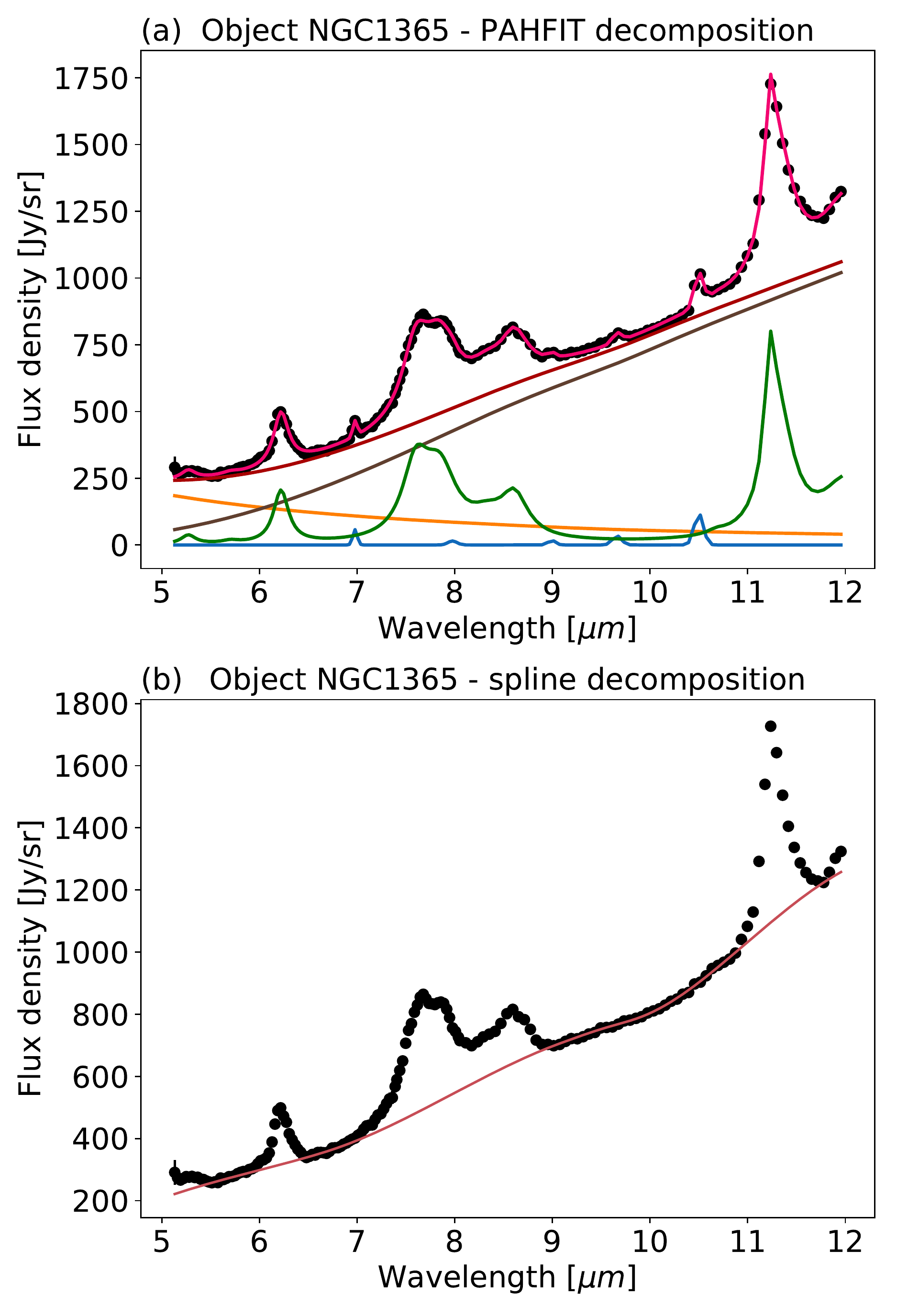}}
\caption{Comparison between the spectral decomposition with PAHFIT and with spline for an object without the PAH plateaus.  (a) Result of the PAHFIT decomposition of the 3C31 spectrum. The data points are represented by the dots with the vertical error-bars as uncertainties.  The pink line corresponds to the best fit model, the green line is the dust (AIB, PAH) contribution and blue one is the ionic and molecular lines contribution. The red line represents the total continuum contribution, and the brown and orange lines correspond to its individual thermal and stellar components, respectively. (b) The red line represents the spline decomposition of the continuum. The data points are represented by the dots with the vertical error-bars as uncertainties.}
\label{fig:pahfit2}
\end{figure}

As already discussed in \citet{Peeters17}, the overall conclusions on PAH intensity correlations for a large sample of objects are independent of the chosen decomposition approach \citep[e.g.][]{Smith07, Galliano08}. However, we analyzed 20 galaxies of our sample -- 10 with strong PAH plateaus and 10 with none or weak plateaus -- in order to perceive the stability of our 6.2~$\mu$m fitting according to the continuum decomposition obtained with spline and with PAHFIT. These galaxies have their 6.2~$\mu$m profile fitted (as it will be discussed in the next section) and the results revealed that the band intensities and the FWHM showed greater discrepancies but the central wavelengths had no significant changes. Therefore, we can conclude that a presence of plateaus did not influence the results. As our work focuses on peak position of the profile, the chosen continuum decomposition applied does not interfere the final analysis. A discussion of this comparison can be found in the Appendix \ref{ap:continuum}. Two examples of these decompositions can be seen in Fig. \ref{fig:pahfit1} and \ref{fig:pahfit2}, with and without PAH plateaus respectively.

\subsection{The 6.2~$\mu$m feature profile}
\label{sub-analysis}

The differences among the 6.2~$\mu$m PAH profiles in such astrophysical environments have been attributed, for example, to the local physical conditions and of the PAH molecules' size, charge, geometry and heterogeneity \citep[e.g][]{Draine01, Draine07, Smith07, sales12}. The CC vibration modes of the 5 -- 9~$\mu$m wavelengths produce profiles (and their peak positions) highly variable, even in relatively low-resolution data \citep{Tielens08}. On the other hand, CH modes vary less and the variation may not be necessarily connected to the others \citep{died04, Tielens08, Candian15}.

According to \citet{hudgins99}, PAHs with roughly 20 carbon atoms already contribute to the 6.2~$\mu$m emission and the spacing among the PAH bands increases with molecular size. In general, for this band, objects grouped in class A have an asymmetric profile composed of a sharp blue rise and a red tail with central wavelength varying up to 6.23~$\mu$m \citep{Peeters02}. In class B and C, this asymmetry decreases and, for the C, the peak position is greater than  6.29~$\mu$m. As already discussed, \citet{Hud05} considered the blueshift of this band (peak positions $\leq$ 6.22~$\mu$m) which characterizes a class A object. The increase of PAH size, the substitution of carbon atoms by silicon or oxygen, the metal ion complexation (Fe$^+$, Mg$^+$ and Mg$^{2+}$) and the molecular symmetry variation were not able to reproduce the observed position of this interstellar band while simultaneously satisfying the astrophysical implications (for instance, the cosmic abundance of the chemically reactive elements). Apparently, of the possible substitutions they consider, nitrogen incorporated into the inner part of the rings is the only solution capable of reproducing the observed profile. The authors also estimated that a lower limit of 1$\%$ -- 2$\%$ of the cosmic nitrogen is retained in the PAH molecules located in the ISM \citep[e.g][]{Hud05, Boersma13, Boersma14a}.

\begin{figure}
\centering
\resizebox{\hsize}{!}{\includegraphics{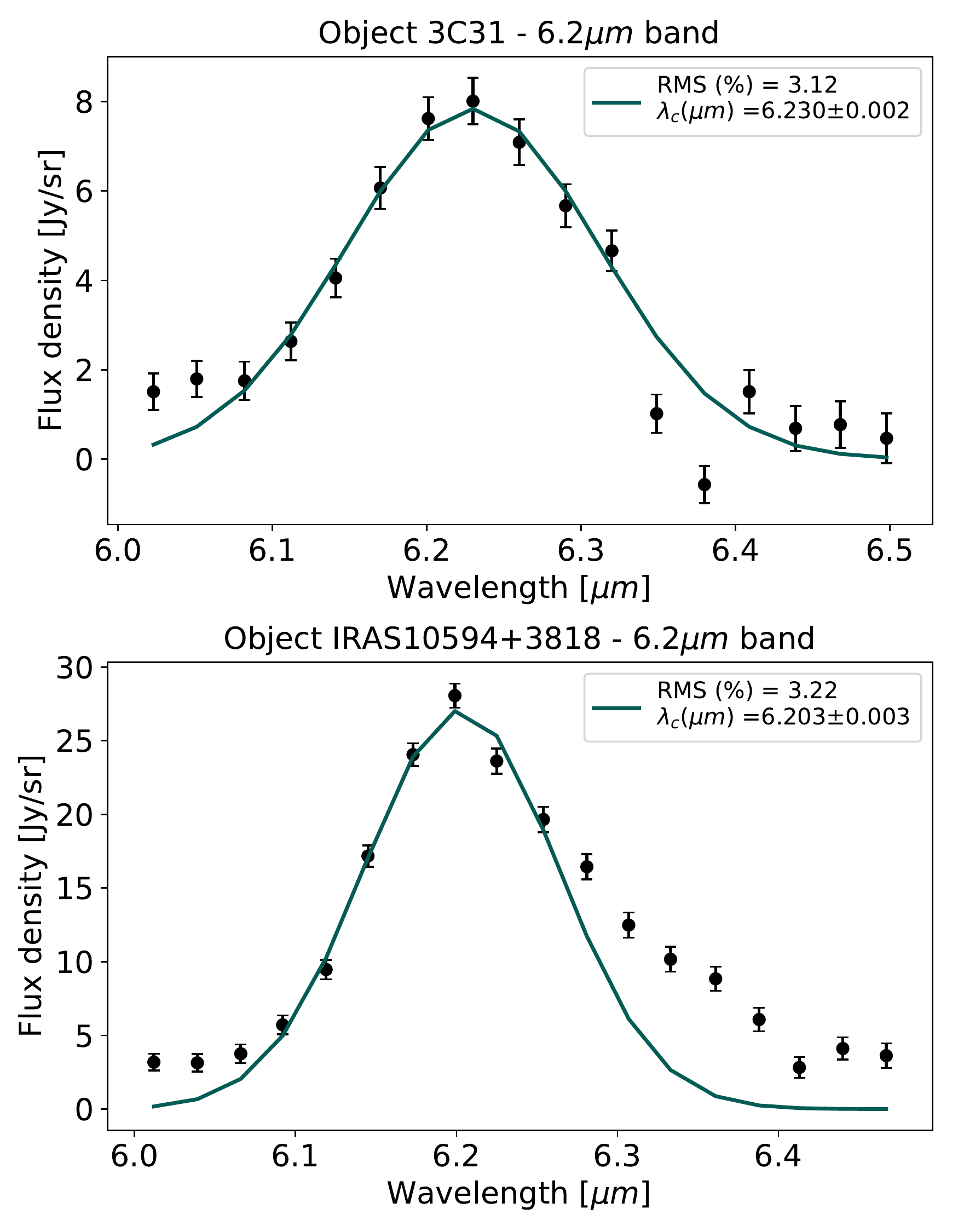}}
\caption{Profile of 6.2~$\mu$m band fitted with \textit{curve\_fit} for the objects 3C31 (\textit{top}) and IRAS10594+3818 (\textit{bottom}). Labels show the values of the peak position and normalized RMS. In the case of IRAS10594+3818, the fitting range was reduced to exclude this more evident red tail.}
\label{fig:cfit}
\end{figure}

To accomplish this study of the 6.2~$\mu$m PAH profile of starburst galaxies, we constructed a python based script to estimate its central wavelength through the optimization algorithms from the submodule \textit{scipy.optmize.curve\_fit} (hereafter \textit{curve\_fit}). The data were fitted by a Gaussian profile (equation~\ref{eq:gauss}). The uncertainties were also derived using this tool with least-squares minimization and we used normalized root mean square (RMS) deviation to evaluate the quality of the fit (Eq. \ref{eq:rmsn}). The initial guesses selected were the values of 6.22~$\mu$m for the peak position and 0.187 for the full width at half maximum (FWHM) \citep{Smith07}. An example of this fit can been seen in Fig.~\ref{fig:cfit} (\textit{top}). 

\begin{equation}
I_{gauss}  = \frac{A}{\sigma \sqrt{2\pi}} exp\left(-\frac{(x-\lambda_c)^2}{2\sigma^2} \right)
\label{eq:gauss}
\end{equation}
where A is the amplitude, $\lambda_c$ is the central wavelength and the full width at half maximum (FWHM) is given by FWHM~$\sim$~2.3548~$\sigma$.

\begin{equation}
RMS (\%)  = \frac{100}{I_{max}} \sqrt{\frac{1}{N} \displaystyle\sum_{i=1}^{N} x_i^2}
\label{eq:rmsn}
\end{equation}
where x$_i^2$ are the quadratic residues, N the number of data and I$_{max}$ is the maximum flux intensity in the evaluated range. RMS values lower than 10\% indicate an appropriate fit and values lower than 5\% indicate a very good fit.

The asymmetry of this specific feature caused the deviation of the fitted peak to redder wavelength and, to handle this, these objects had the data points of the profile's red tail removed from the fit following the same method of \citet{Peeters17}. One possible contributor to this red tail is anharmonic hot-band emission which is a natural consequence of the PAH model, although it is not a major player in determining the profile itself \citep{hudgins99}. Some galaxies of our sample required a reduction in the fitting range to better exclude the red tail and they are marked with an asterisk in Table~\ref{tab:cfit}. An example of this case is shown in Fig. \ref{fig:cfit} (\textit{bottom}).

\begin{figure}
\centering
\resizebox{\hsize}{!}{\includegraphics{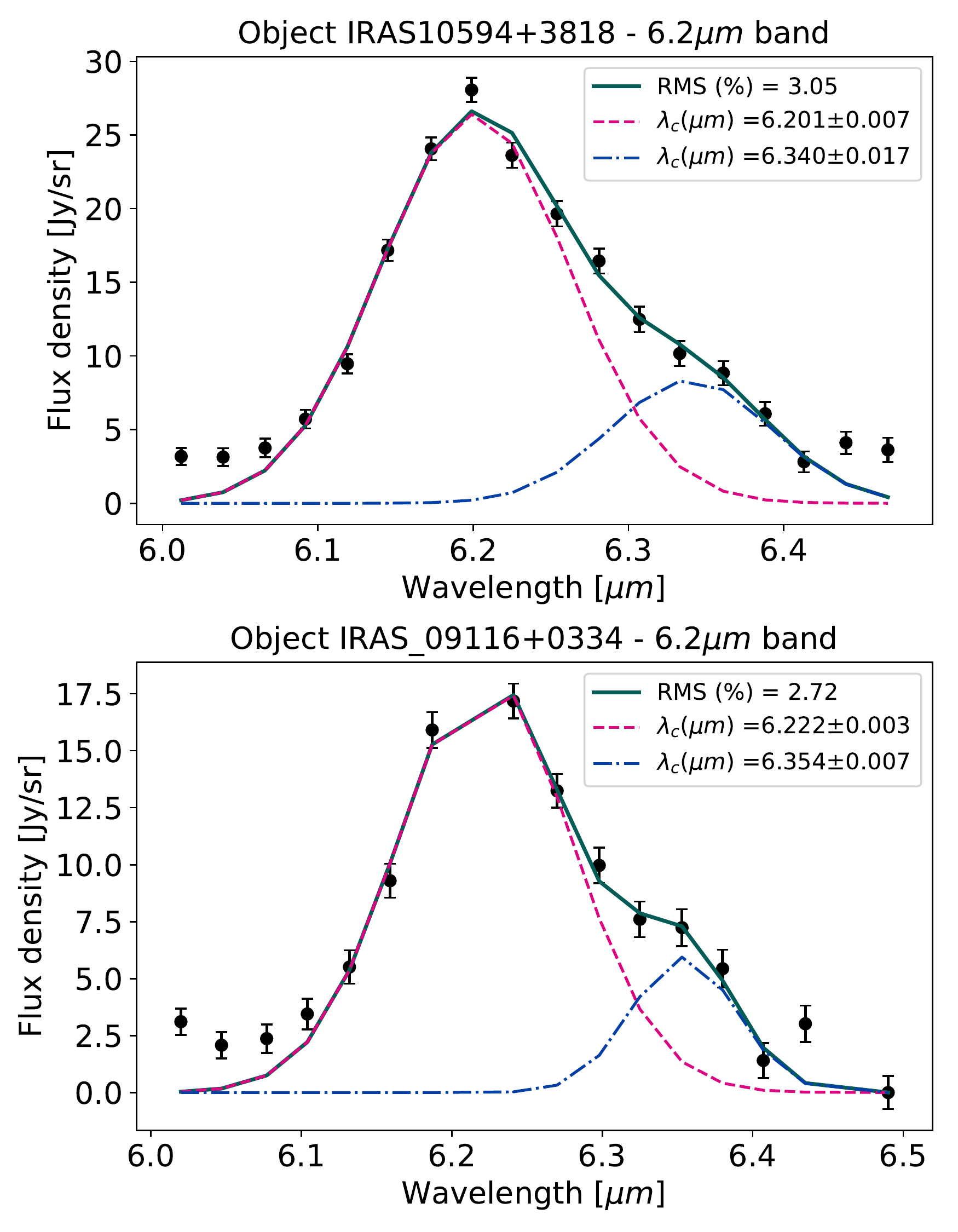}}
\caption{Profile of 6.2~$\mu$m band fitted with \textit{curve\_fit} for the objects IRAS10594+3818 (\textit{top}) and IRAS\_09116+0334 (\textit{bottom}) using two different Gaussian profiles. Labels show the values of the peak positions and normalized RMS.}
\label{fig:cfit2}
\end{figure}

We also noted the presence of another emission component of the 6.2~$\mu$m band near 6.35~$\mu$m  in some sources and performed the fitting adding a new subcomponent to the 6.22~$\mu$m band (Fig. \ref{fig:cfit2}, \textit{top}) in order to compare the two procedures. The inclusion of the second feature seems to encompass the fit for highly asymmetric profiles by reproducing their red tail, previously underestimated in some galaxies, as for the object IRAS10594+3818 (Fig. \ref{fig:cfit2}, \textit{bottom}). Even so, in these situations, there is no real indication of the presence of the second feature and the asymmetry could be just a characteristic of the anharmonic profile \citep{Tielens08}. In this work, we used only the fitting with just one Gaussian to standardize the analysis.

With peak positions obtained, we were able to group the galaxies into the three Peeters' classes (Table~\ref{tab:classes}). Fig.~\ref{fig:zXwl-all} show the class distribution.

\begin{table}
	\centering
	\caption{Intervals for each Peeters' classes \citep{Peeters02}.}
	\label{tab:classes}
	\begin{tabular}{cc}
		\hline
		Class & Interval ($\mu m$)\\
		\hline
		A & $<6.23$\\
		B & $6.23< \lambda <6.29$\\
		C & $>6.29$\\
		\hline
	\end{tabular}
\end{table}

\section{Results and Discussion}
\label{results}

The second feature near 6.35~$\mu$m has been attributed in some cases to the inherent asymmetry of the band profile. Nevertheless, its peak position varied from 6.246 to 6.471~$\mu$m and there are a few possibilities that could give insight into its nature. According to \citet{pino08}, who performed experiments with PAHs in order to verify the origins of their emissions, bands near 6.3~$\mu$m may be related to aliphatic features. On the other hand, the  6.4~$\mu$m band observed in the reflection nebula NGC7023 was attributed to the C$^{+}_{60}$ \citep{berne15}. PAH cations, without any carbon substituted, could also be the responsible for this emission \citep{Hud05}. Finally, peaks at $\approx$ 6.41~$\mu$m could be due to perylene-like structures \citep{Candian14}.
	 
\begin{figure}
\centering
\resizebox{\hsize}{!}{\includegraphics{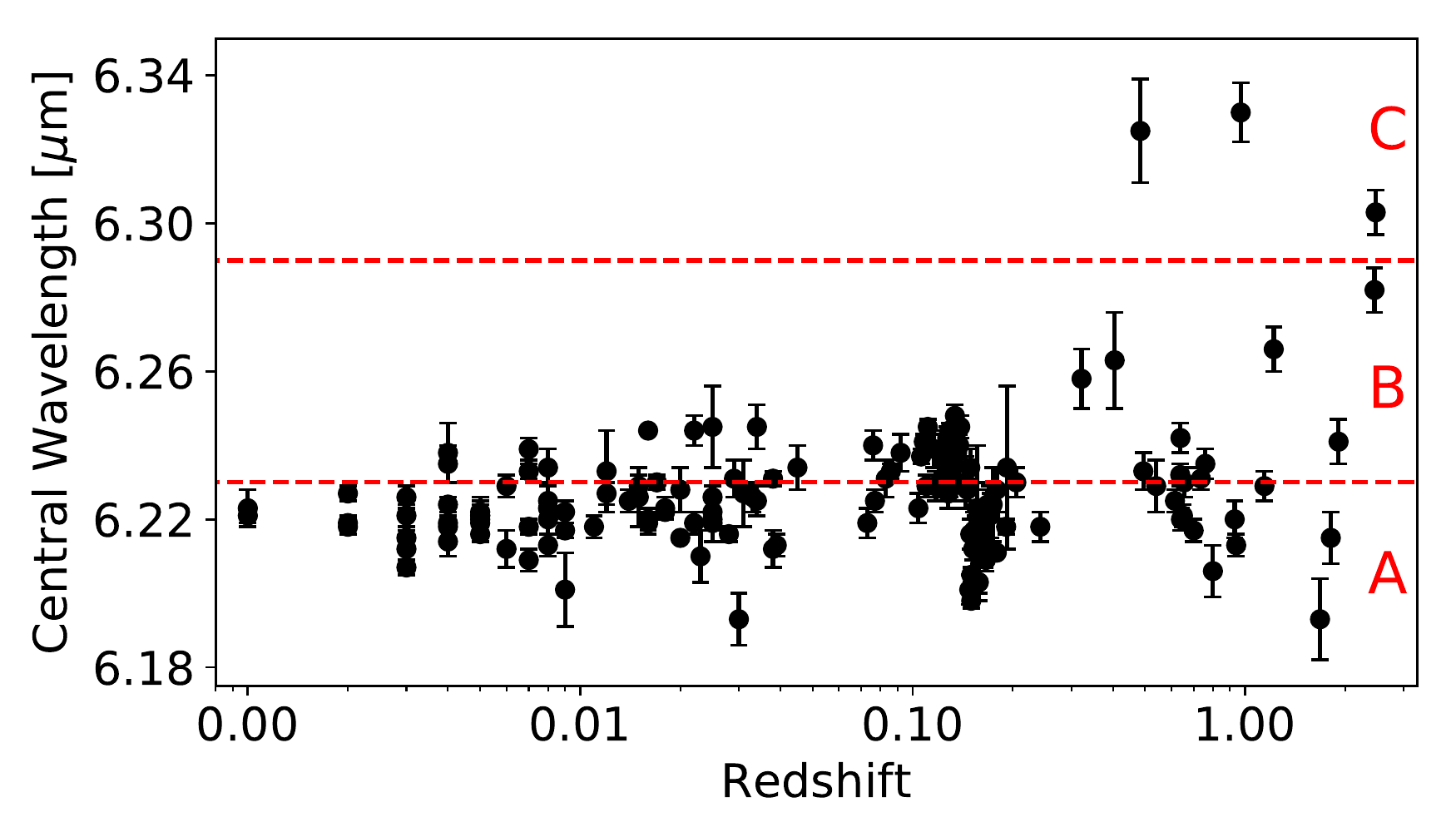}}
\caption{Distribution of the central wavelengths of the 6.2~$\mu$m band for the redshift of the sources. The x axis is in logarithmic scale. The dashed lines are the limits among the Peeters' classes, represented also by the letters A, B and C.}
\label{fig:zXwl-all}
\end{figure}

Table~\ref{tab:results-classes} summarizes the separation of the 155 objects into Peeters' classes derived from the \textit{curve\_fit} fits. Details are given in Table~\ref{tab:cfit}.  An overview of the results is shown in Fig.~\ref{fig:zXwl-all}, which displays the distribution of the central wavelengths taking into account of redshifts of the galaxies. There is a small predominance of class A objects over class B objects (more evident for redshifts lower than 0.05). Only three galaxies were classified as C. 

\begin{table}
	\centering
	\caption{Number of galaxies that fall into each Peeters' class.}
	\label{tab:results-classes}
	\begin{tabular}{cccc}
		\hline
		\textbf{Galaxies} & \textbf{Class A} & \textbf{Class B} & \textbf{Class C}\\
		\hline
		155 & 67\% & 31\% & 2\%  \\
		\hline
	\end{tabular}
\end{table}

\citet{pino08} have already noticed that class A objects are the most common in the Universe and embrace several astrophysical sources, while class C objects are in the minority. Our results point to the same conclusions, especially for starburst-dominated sources whose class A members are up to 67\% in our study. This evidence is more pronounced at lower redshifts, as can be seem in Fig.~\ref{fig:zXwl-all}. In addition, if we consider class B as a mixture between PANH and PAH emissions, as pointed out by \citet{Peeters02}, we can verify that PANHs dominate this sub-class of galaxies based on analysis of the 6.2~$\mu$m PAH.

\begin{figure}
\centering
\resizebox{\hsize}{!}{\includegraphics{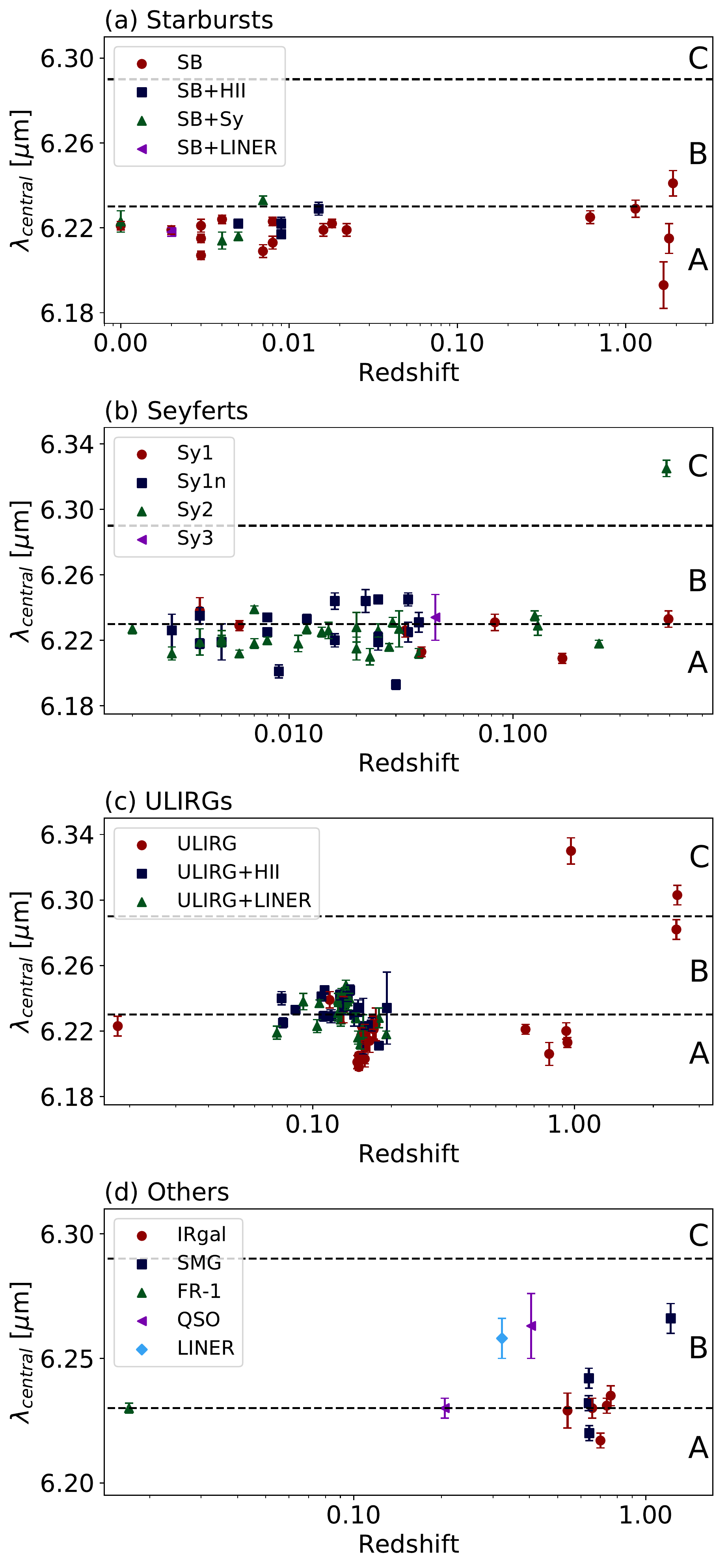}}
\caption{Distribution of the central wavelengths of the 6.2~$\mu$m band according to the galaxy type -- (a) Starbursts; (b) Seyferts; (c) ULIRGs; and (d) Others. The dashed lines are the limits among the Peeters' classes, indicated also by A, B or C letter. Acronyms: AGN -- Active Galactic Nucleus, FR -- Fanaroff-Riley galaxy, HII --  HII region, IRgal -- Infrared galaxy, LINER -- Low-Ionization Nuclear Emission-line Region, QSO -- Quasi-Stellar Object, SB -- Starburst galaxy, SMG -- Submillimeter Galaxy, Sy -- Seyfert galaxy, ULIRG -- Ultra-Luminous Infrared Galaxy.}
\label{fig:tot1}
\end{figure}

Regarding class C, just three objects do not allow us to distinguish any pattern. Apparently, they may be expected in higher redshifts, which might also imply an evolutionary timescale of PAH molecules. Chemically young astrophysical sources might have reduced PAH abundances and PAH molecules are not as efficiently produced in low-metallicity environments because fewer carbon atoms are available in the ISM \citep{Shivaei17}. In this case, class C could be represented by VSGs (very small grains). It was already noticed that VSGs may be responsible for the extended red wing and redshift of the peak position of the 11.2~$\mu$m PAH band \citep{Rosenberg11}. Besides, they may also be the carrier of the 7.8~$\mu$m subcomponent of the 7.7~$\mu$m PAH band and the 8.25~$\mu$m component of the 8.6~$\mu$m PAH band \citep{Peeters17}, which are expected to be stronger for class C objects (Section \ref{intro}). Nevertheless, as the higher redshift objects were extrapolated by the PAH bands themselves and were not corroborated with other spectroscopic or photometric data, any additional analysis may be misguided because the redshift errors can be as large as 0.1 -- 0.2 depending on how many features were used in the calculation \citep{yan07}.

According to \citet{ota16}, the substitution of three or more nitrogen atoms into PAHs does not provide molecules that describe the observed features. This could indicate that compounds of astrobiological interest such as purine and adenine may not be synthesized in the ISM. On the other hand, considering just one or two nitrogens, some species were capable of reproducing the observations (C$_7$H$_5$N$_2$-ab$^{3+}$, for example). However, only small PANHs were considered in their study and they are easily destroyed in the ISM than larger molecules. We can expect that  larger PANHs are correlated to the 6.2~$\mu$m band emission since PAHs with $>$~20 -- 30 carbon atoms are thought to dominate the emitting interstellar PAHs.

\begin{table}
	\centering
	\caption{Peeters' class distribution for starbursts, Seyferts and ULIRG galaxies.}
	\label{tab:classes-kinds}
	\begin{tabular}{ccccc}
		\hline \
		\textbf{Object} & \textbf{A} &  \textbf{B} &   \textbf{C} & \textbf{Total of the}\\
		& \textbf{(\%)} &  \textbf{(\%)} &   \textbf{(\%)} & \textbf{sample (\%)}\\
		\hline
		SB & 92 & 8 & 0 & 17\\
		SB + HII &  &  & &  \\
		SB + Sy &  &  & &  \\
		\hline
		Sy1 & 70 & 28 & 2 & 34\\
		Sy1n &  &  & &  \\
		Sy2 &  &  &  & \\
		Sy3 &  &  &  & \\		
		\hline 
		ULIRG & 59 & 38 & 3 & 41\\
		ULIRG + HII &  &  &  & \\
		ULIRG + LINER &  &  &  &\\ 
		\hline
		Others & 46 & 54 & 0 & 8\\
		\hline
	\end{tabular}
\end{table}

Fig.~\ref{fig:tot1} illustrates the class distribution in different galaxy types for both methods applied (see Table \ref{tab:sources}). Table~\ref{tab:classes-kinds} presents the percentages of each class for the three most abundant objects of the galaxies -- starbursts (17\%), Seyferts (34\%) and ULIRGs (41\%). The other types comprehend 8\%.

In the scenario in which class A 6.2~$\mu$m band position arises from PANHs, the importance of these species is greater in starbursts. However, the fact that most of the ULIRGs and Seyferts are class A objects indicates a significant presence of PANHs also in these environments. The prevalence of class A objects in our study could be explained by the dominance of star formation contribution in all sources of our sample, or by the ubiquity of PANHs in galaxies. Studies of samples with a larger number of AGN-dominated sources could help to clarify this issue. Only in SMGs, class B dominates: 3 in 4 galaxies are classified as B. Again, more data is needed for further analysis.

\begin{figure}
\centering
\resizebox{\hsize}{!}{\includegraphics{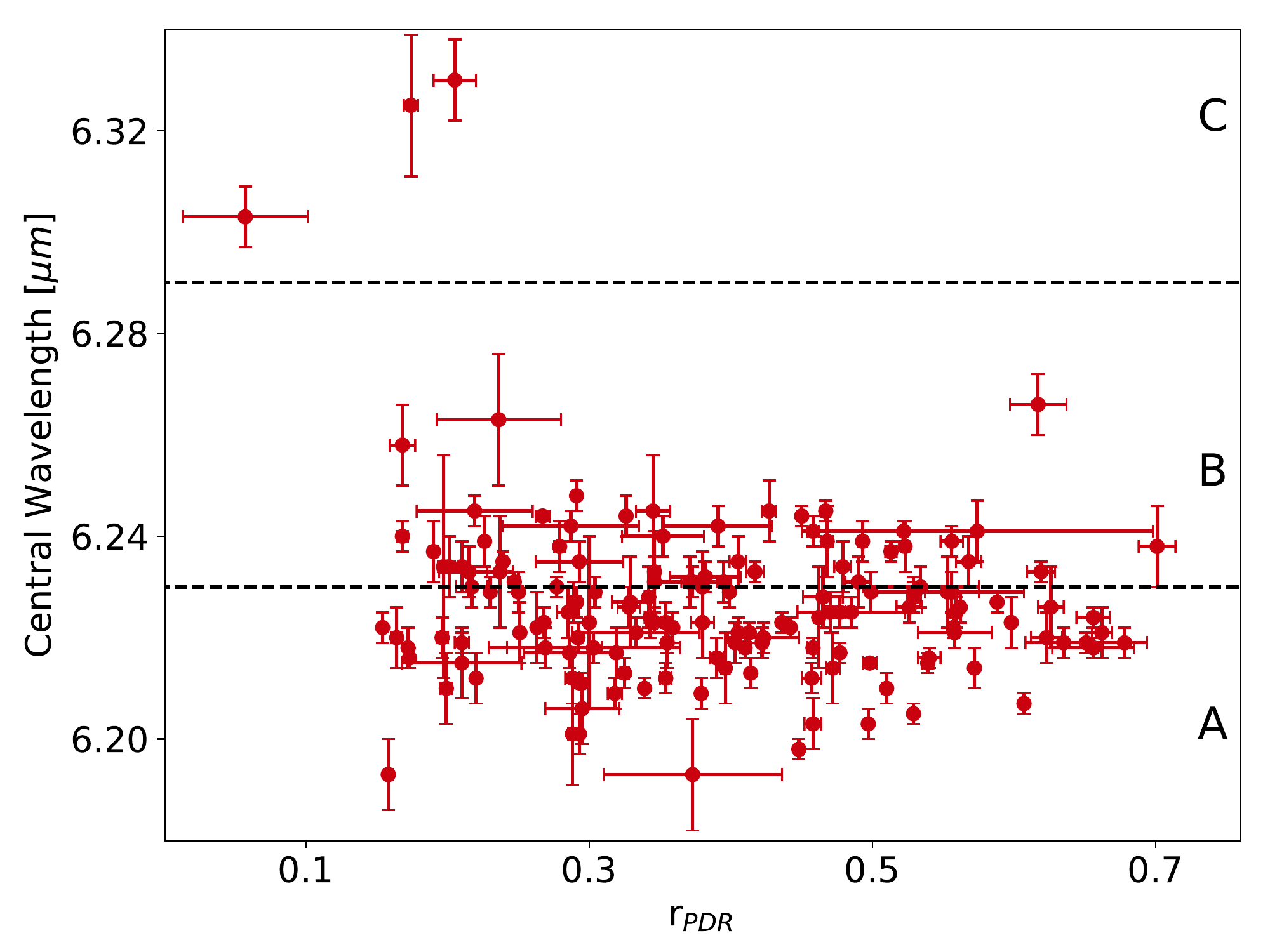}}
\caption{Distribution of the central wavelengths of the 6.2~$\mu$m band for the spectral contribution of PDR component r$_{PDR}$ of 152 sources which values were available in the ATLAS project. The uncertainties are displayed as errorbars when available.The dashed lines are the limits among the Peeters' classes, represented also by letters A, B and C.}
\label{fig:rpdr}
\end{figure}

Fig.~\ref{fig:rpdr} shows the distribution of the central wavelengths of the 6.2~$\mu$m band for the spectral contribution of the PDR component (r$_{PDR}$) of the sources, as calculated by \citet{caballero}. The r$_{PDR}$ value is the ratio of the total integrated luminosity to the total luminosity of the PDR component in the 5 -- 15~$\mu$m rest-frame range. In this sense, the class C sources present low contribution of star formation and greater contribution of the ANG or HII region components, also calculated by \citet{caballero}. Table \ref{tab:Ccomp} shows these respective values for the three class C sources.

\begin{table}
	\centering
	\caption{Spectral contribution of the AGN, HII and PDR components for the class C objects \citep{caballero}.}
	\label{tab:Ccomp}
	\begin{tabular}{lccc}
		\hline \
		\textbf{Object} & \textbf{r$_{AGN}$} &  \textbf{r$_{HII}$} &   \textbf{r$_{PDR}$} \\
		\hline 
		MIPS 180 & 0.439$\pm$0.075 & 0.504$\pm$0.112 & 0.057$\pm$0.044\\
		\hline 
		SDSS\_J00562 & 0.571$\pm$0.008 & 0.256$\pm$0.006 & 0.174$\pm$0.005\\ 
		1.72+003235.8 & & \\
		\hline 		
        SWIRE4\_J1036 & 0.458$\pm$0.022 & 0.337$\pm$0.014 & 0.205$\pm$0.015\\
        37.18+584217.0 & & \\
        \hline 
	\end{tabular}
\end{table} 

\begin{figure}
\centering
\resizebox{\hsize}{!}{\includegraphics{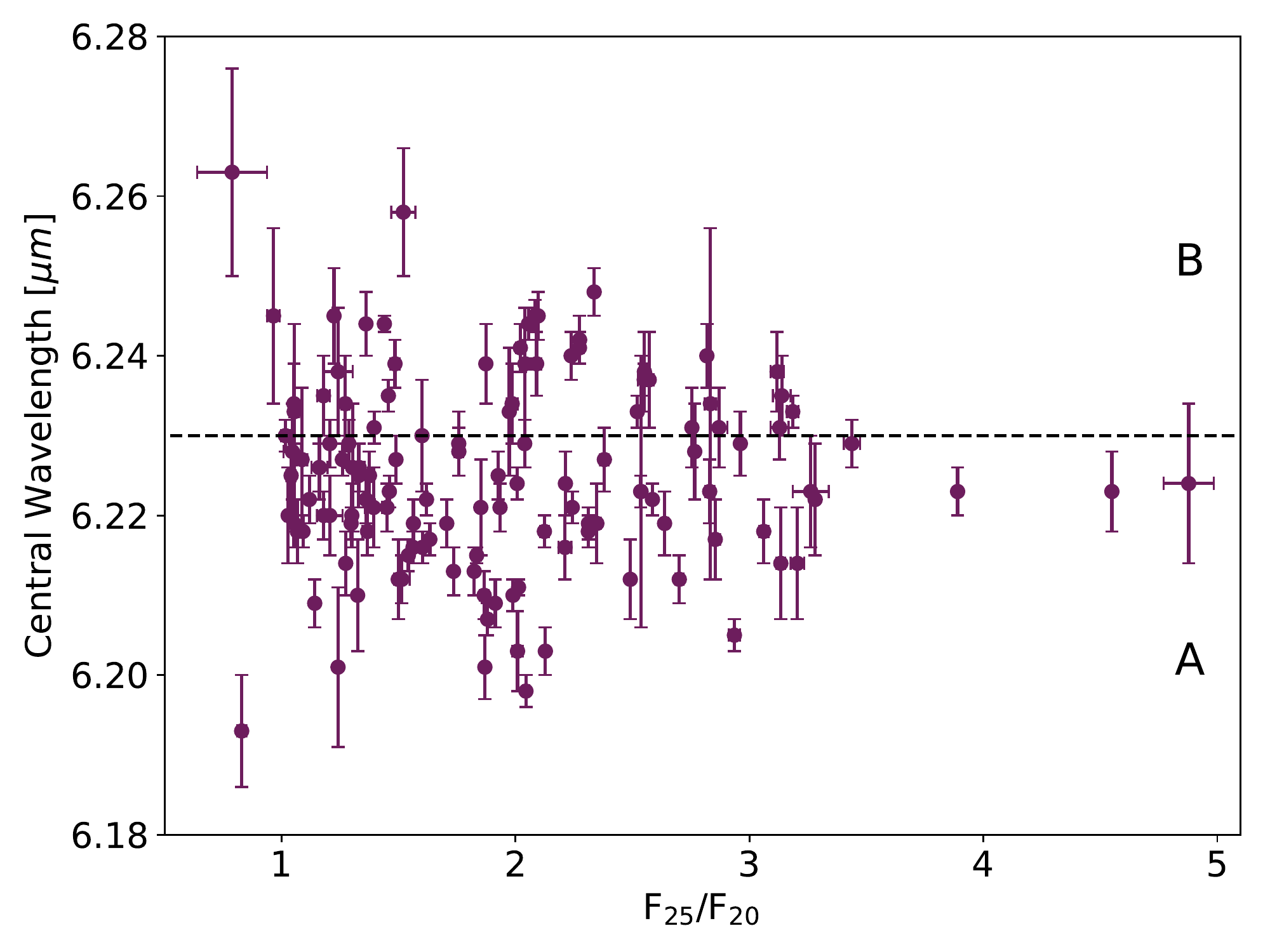}}
\caption{Distribution of the central wavelengths of the 6.2~$\mu$m band for the F$_{25}$/F$_{20}$ ratio of 129 sources which values were available in the ATLAS project. The uncertainties are displayed as errorbars when available. The dashed lines are the limits among the Peeters' classes, represented also by letters A, B and C.}
\label{fig:f25-20}
\end{figure}

On the other hand, if we consider the F$_{25}$/F$_{20}$ ratio (Fig.~\ref{fig:f25-20}), we can see that class A objects are the coldest. The values of r$_{PDR}$, F$_{25}$ and F$_{20}$ together with their respective uncertainties were extracted directly from the ATLAS project when they were available. The uncertainty of the F$_{25}$/F$_{20}$ ratio was propagated from the F$_{25}$ and F$_{20}$ errors. The values of r$_{PDR}$ and F$_{25}$/F$_{20}$ ratio can be seen in Table \ref{tab:Tidx}.

\section{Conclusions}
\label{conclusion}
We have analyzed the MIR spectra of 155 starburst-dominated galaxies, searching for the contribution of the Peeters' class A 6.2$\mu$m band to the total sample. To date, the class A position of this band can only be attributed to PANHs, PAHs containing N atoms. Thus, the PAH feature profiles, especially the 6.2~$\mu$m band, could indicate the presence of nitrogen incorporated to the rings. 

The fitted 6.2~$\mu$m profiles were classified in classes A, B and C following \citet{Peeters02}. At 67\% of the sample, class A profiles clearly dominate, suggesting a significant presence of PANHs in the ISM of these galaxies. Class B corresponded to a percentage of 31\%, indicating a significantly smaller contribution of PANHs in these sources. Only class C, with a small percentage of 2\%, seems not to be influenced by these molecules. These trends give support to the suggestion that class A/B 6.2~$\mu$m band variations track changes in PANH ionization state or molecular structure \citep{Baus09}. In addition, we can see that class A objects are colder compared to class B objects.

Within the PANH scenario, the ubiquity of PANHs could indicate another reservoir of nitrogen in the Universe, with density and temperature conditions that differ from those of gas and ices phases. As shown in Fig.~\ref{fig:tot1}, they can be present in the ISM of starburts galaxies, ULIRGs, Seyferts, infrared and submillimeter galaxies. Furthermore, they are responsible for an important fraction of the MIR emission, especially for the 6.2~$\mu$m band. These findings also give support to the idea of their contribution to the origins of life on Earth and elsewhere, since they could form nucleobase-type structures in the ISM \citep{Elsila06, Parker15a}.

Extension of this analysis to other types of objects also available in the ATLAS project, such as AGNs, could shed light on how the starburst-dominated emission of the sources is responsible for the majority of class A objects and could provide a broader overview of the 6.2~$\mu$m band behavior in astrophysical environments. 

Moreover, it will also be possible to explore the other PAH bands in more detail, which could be of great interest to this study. Since 6.2 and 7.7~$\mu$m bands are both caused by the CC stretching vibrational mode, they are connected to each other in some cases, mainly for class A \citep{died04}. This association could furnish another strategy for deriving the variations of the 6.2~$\mu$m band in an indirect way. 

The question why the presence of PANHs is more apparent in some galaxies but not in others could be addressed with chemical evolution models taking into account differences in metallicity, star formation history and the nature of molecular clouds in the harboring galaxy, e.g. the chemodynamical model in \citet{Friaca17}. In addition, further computational calculations together with laboratory measurements are needed to make more robust predictions of the role of PANHs in the profile of the PAH emission bands, mainly in the conditions prevailing in galaxies with star formation.

\section*{Acknowledgements}

Special acknowledgements to CAPES (Comiss\~ao de Aperfei\c{c}oamento de Pessoal do N\'ivel Superior) and CNPq (Conselho Nacional de Desenvolvimento Cient\'ifico e Tecnol\'ogico) for the financial support.




\bibliographystyle{mnras}
\bibliography{bib17} 




\appendix
\onecolumn

\section{Sources -- identification and derived properties}

\input{table_sources.tex}
\input{table_62curvefit.tex}

\input{table_Tidx.tex}

\twocolumn
\section{}
\label{ap:continuum}
\subsection*{Comparison of the 6.2~$\mu$m results for the continuum decomposition}

\begin{figure}
\centering
\includegraphics[scale=0.43]{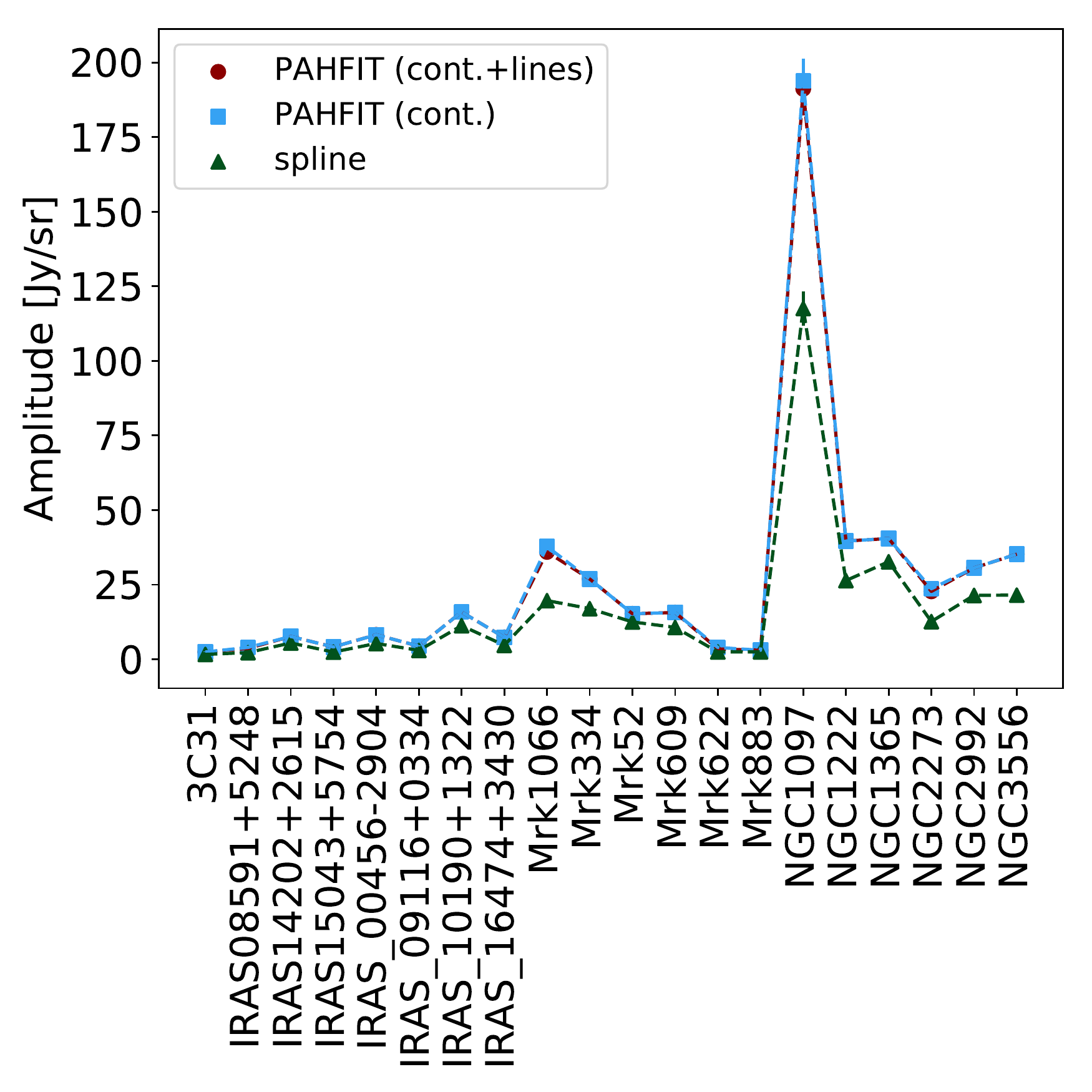} 
\caption{Results for the 6.2$\mu$m profile amplitude depending on each decomposition method -- (i) circle; (ii) square; and (iii) triangle.}
\label{fig:comp-amp}
\end{figure}

\begin{figure}
\centering
\includegraphics[scale=0.43]{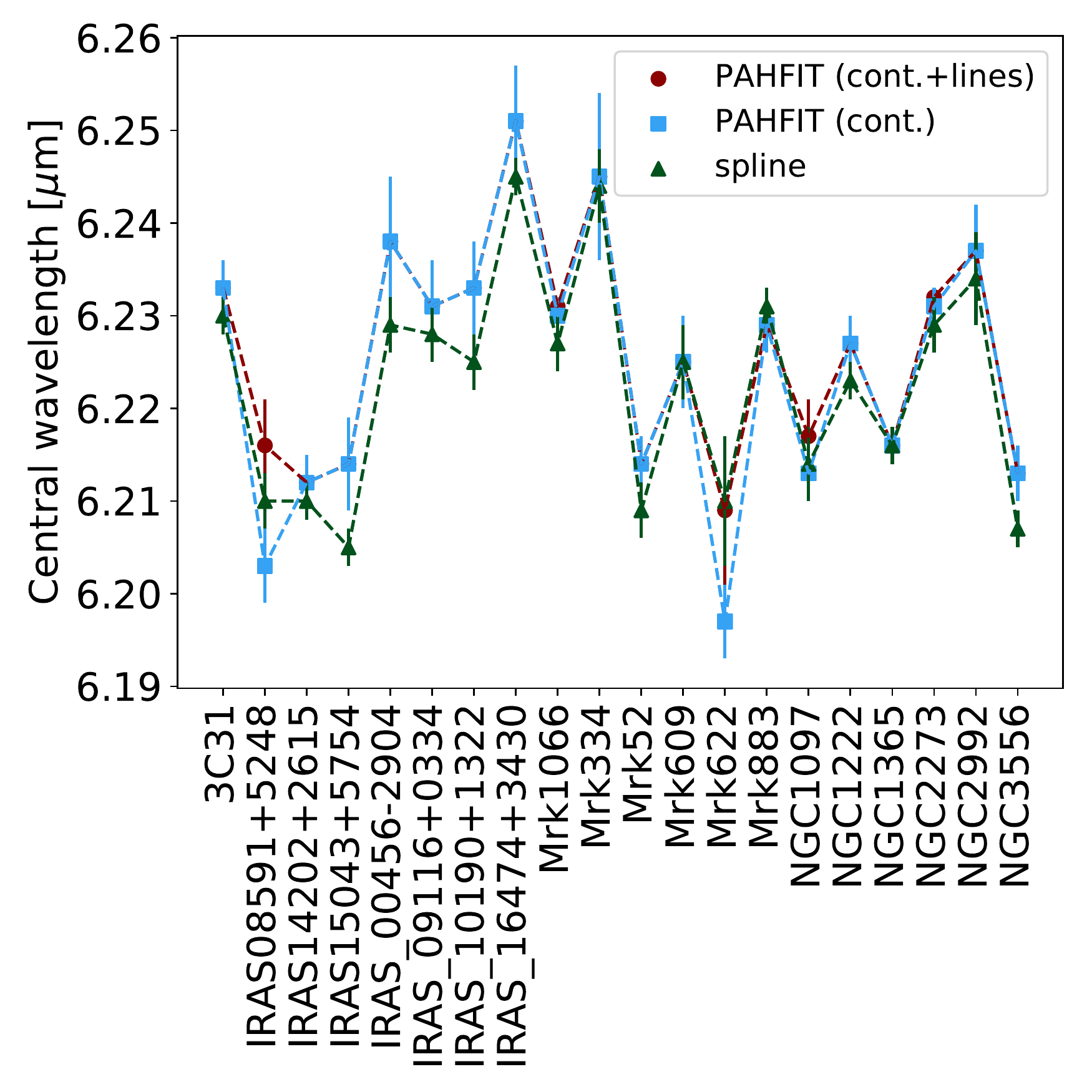} 
\caption{Results for the 6.2$\mu$m profile central wavelength depending on each decomposition method -- (i) circle; (ii) square; and (iii) triangle.}
\label{fig:comp-wl}
\end{figure}

\begin{figure}
\centering
\includegraphics[scale=0.43]{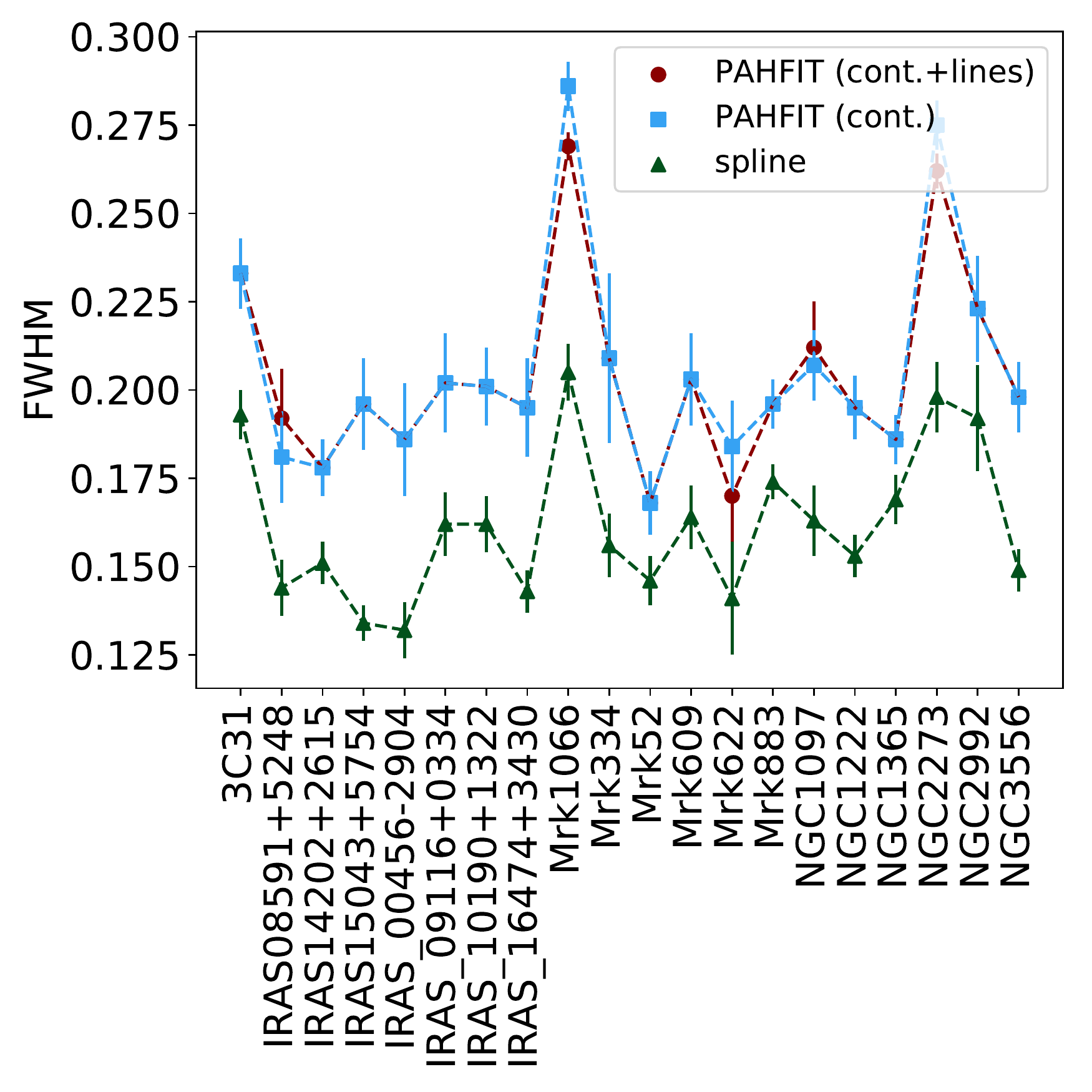} 
\caption{Results for the 6.2$\mu$m profile FWHM depending on each decomposition method -- (i) circle; (ii) square; and (iii) triangle.}
\label{fig:comp-fwhm}
\end{figure}

The Tables \ref{tab:cont-amp}, \ref{tab:cont-wl} and \ref{tab:cont-fwhm} and the Figures \ref{fig:comp-amp}, \ref{fig:comp-wl} and \ref{fig:comp-fwhm} compare the differences among the 6.2~$\mu$m band fits according to the spectral continuum contribution fitted with the three methods described bellow. Such contribution was subtracted from the spectra before the band was fitted. 20 galaxies were selected for this comparison, 10 with strong PAH plateaus and 10 with weak or none plateaus (see section \ref{analysis}). To standardize the comparison, in all cases the band was fitted using the same method. We fitted using only one Gaussian profile and the submodule \textit{curve\_fit}. The following methodology was applied.

\begin{description}
\item (i) subtraction of the continuum and ionic and molecular contributions fitted with PAHFIT 
\item (ii) only the subtraction of the continuum contribution fitted with PAHFIT
\item (iii) subtraction of the ionic and molecular contributions fitted with PAHFIT and the subtraction of the continuum contribution decomposed with spline
\end{description} 

The central wavelengths, which are the objective of this work, presented values range within the uncertainty bar of the parameter. Therefore, the method chosen does not influence the final results in our analysis. On the other hand, the band intensity and the FWHM even doubled their values depending on whether the fit of the continuum was performed by PAHFIT or by the spline.

The methods (i) and (ii) resulted, in general, in the same values. The only difference between both is whether the ionic and molecular lines contributions were subtracted or not from the spectra before the band analysis. As no large blending of these lines is known, this similarity in the results was already expected, revealing that the contribution of these lines is insignificant for the final results.

Therefore, the greatest discrepancy is due to the program used. To better understand the results, the spline decomposition is preferable to the PAHFIT fit for our particular analysis because it already considers the PAH plateau of 5--10~$\mu$m rather than diluting it in the Drude profiles of the PAH bands \citep{Peeters17}. Moreover, this difference in the band intensity is crucial when considering other PAH bands such as the 7.7~$\mu$m band, for instance.

\onecolumn

\input{tables_continua.tex}


\bsp	
\label{lastpage}
\end{document}

%% file: table_sources.tex
\topcaption{Sources and their respective informations extracted from the MIR\_SB sample (Spitzer/IRS ATLAS, version 1.0) and \citet{yan07}, including their ID, type, source reference, right ascension, declination and redshift. Acronyms: AGN -- Active Galactic Nucleus, FR -- Fanaroff-Riley galaxy, HII --  HII region, IRgal -- Infrared galaxy, LINER -- Low-Ionization Nuclear Emission-line Region, QSO -- Quasi-Stellar Object, SB -- Starburst galaxy, SMG -- Submillimeter Galaxy, Sy -- Seyfert galaxy, ULIRG -- Ultra-Luminous Infrared Galaxy.}\label{tab:sources}

\tablefirsthead{\toprule 
ID &
\multicolumn{1}{c}{Type} &
\multicolumn{1}{c}{Reference} &
\multicolumn{1}{c}{RA (hms)} &
\multicolumn{1}{c}{Dec (dms)} &
\multicolumn{1}{c}{z}

 \\ \midrule}

\tablehead{%
\multicolumn{6}{c}%
{{\bfseries  Continued from previous page}} \\
\toprule
ID &
\multicolumn{1}{c}{Type} &
\multicolumn{1}{c}{Reference} &
\multicolumn{1}{c}{RA (hms)} &
\multicolumn{1}{c}{Dec (dms)} &
\multicolumn{1}{c}{z}  \\ \midrule}
\tabletail{%
\midrule \multicolumn{6}{r}{{Continued on next page}} \\ \midrule}
\tablelasttail{%
\\\midrule
\multicolumn{6}{r}{{$^a$Objects with redshift obtained through the IRS spectrum.}} }

\begin{center}
\begin{xtabular}{lccccc}

3C293 & Sy3 & \citet{Leipski09} & 13:52:17.80 & 31:26:46.50 & 0.045 \\
3C31 & FR-1 & \citet{Leipski09} & 01:07:24.90 & 32:24:45.20 & 0.017 \\
AGN15$^a$ & LINER & \citet{Weedman09} & 17:18:52.71 & 59:14:32.00 & 0.322 \\
Arp220 & ULIRG & \citet{Imanishi07} & 15:34:57.10 & 23:30:11.00 & 0.018 \\
CGCG381-051 & Sy2 & \citet{Wu09} & 23:48:41.70 & 02:14:23.00 & 0.031 \\
E12-G21 & Sy1 & \citet{Wu09} & 00:40:47.80 & -79:14:27.00 & 0.033 \\
EIRS-2$^a$ & SB & \citet{Hernot-Caballero09} & 16:13:49.94 & 54:26:28.40 & 1.143 \\
EIRS-14$^a$ & SB & \citet{Hernot-Caballero09} & 16:35:36.64 & 40:47:53.80 & 0.615 \\
EIRS-41 & QSO & \citet{Hernot-Caballero09} & 16:34:28.15 & 41:27:42.60 & 0.405 \\
GN26 & SMG & \citet{Pope08} & 12:36:34.51 & 62:12:40.90 & 1.219 \\
IC342 & SB & \citet{Brandl06} & 03:46:48.51 & 68:05:46.00 & 0.001 \\
IRAS02021-2103 & ULIRG & \citet{Imanishi10} & 02:04:27.30 & -20:49:41 & 0.116 \\
IRAS02480-3745 & ULIRG & \citet{Imanishi10} & 02:50:01.70 & -37:32:45 & 0.165 \\
IRAS03209-0806 & Sy1 & \citet{Imanishi10} & 03:23:22.90 & -07:56:15 & 0.166 \\
IRAS04074-2801 & ULIRG & \citet{Imanishi10} & 04:09:30.40 & -27:53:44 & 0.153 \\
IRAS05020-2941 & ULIRG & \citet{Imanishi10} & 05:04:00.70 & -29:36:55 & 0.154 \\
IRAS08591+5248 & ULIRG & \citet{Imanishi10} & 09:02:47.50 & 52:36:30 & 0.158 \\
IRAS10594+3818 & ULIRG & \citet{Imanishi10} & 11:02:14.00 & 38:02:35 & 0.158 \\
IRAS12447+3721 & ULIRG & \citet{Imanishi10} & 12:47:07.70 & 37:05:37 & 0.158 \\
IRAS13106-0922 & ULIRG & \citet{Imanishi10} & 13:13:14.80 & -09:38:00 & 0.174 \\
IRAS14121-0126 & ULIRG & \citet{Imanishi10} & 14:14:45.50 & -01:40:55 & 0.150 \\
IRAS14197+0813 & ULIRG & \citet{Imanishi10} & 14:22:11.60 & 07:59:28AM & 0.131 \\
IRAS14202+2615 & ULIRG & \citet{Imanishi10} & 14:22:31.40 & 26:02:05 & 0.159 \\
IRAS14485-2434 & ULIRG & \citet{Imanishi10} & 14:51:23.80 & -24:46:30 & 0.148 \\
IRAS15043+5754 & ULIRG & \citet{Imanishi10} & 15:05:39.50 & 57:43:07 & 0.150 \\
IRAS21477+0502 & ULIRG & \citet{Imanishi10} & 21:50:16.30 & 05:16:03AM & 0.171 \\
IRAS22088-1831 & ULIRG & \citet{Imanishi10} & 22:11:33.80 & -18:17:06 & 0.170 \\
IRAS\_00091-0738 & ULIRG + HII & \citet{Imanishi07} & 00:11:43.30 & -07:22:07.00 & 0.118 \\
IRAS\_00456-2904 & ULIRG + HII & \citet{Imanishi07} & 00:48:06.80 & -28:48:19.00 & 0.110 \\
IRAS\_01199-2307 & ULIRG + HII & \citet{Imanishi09} & 01:22:20.90 & -22:52:07.00 & 0.156 \\
IRAS\_01355-1814 & ULIRG + HII & \citet{Imanishi09} & 01:37:57.40 & -17:59:21.00 & 0.192 \\
IRAS\_01494-1845 & ULIRG & \citet{Imanishi09} & 01:51:51.40 & -18:30:46.00 & 0.158 \\
IRAS\_02411+0353 & ULIRG + HII & \citet{Imanishi07} & 02:43:46.10 & 04:06:37.00 & 0.144 \\
IRAS\_03250+1606 & ULIRG + LINER & \citet{Imanishi07} & 03:27:49.80 & 16:17:00.00 & 0.129 \\
IRAS\_03521+0028 & ULIRG + LINER & \citet{Imanishi09} & 03:54:42.20 & 00:37:03.00 & 0.152 \\
IRAS\_08201+2801 & ULIRG + HII & \citet{Imanishi09} & 08:23:12.60 & 27:51:40.00 & 0.168 \\
IRAS\_09039+0503 & ULIRG + LINER & \citet{Imanishi07} & 09:06:34.20 & 04:51:25.00 & 0.125 \\
IRAS\_09116+0334 & ULIRG + LINER & \citet{Imanishi07} & 09:14:13.80 & 03:22:01.00 & 0.146 \\
IRAS\_09463+8141 & ULIRG + LINER & \citet{Imanishi09} & 09:53:00.50 & 81:27:28.00 & 0.156 \\
IRAS\_09539+0857 & Sy2 & \citet{Imanishi07} & 09:56:34.30 & 08:43:06.00 & 0.129 \\
IRAS\_10190+1322 & ULIRG + HII & \citet{Imanishi07} & 10:21:42.50 & 13:06:54.00 & 0.077 \\
IRAS\_10485-1447 & ULIRG + LINER & \citet{Imanishi07} & 10:51:03.10 & -15:03:22.00 & 0.133 \\
IRAS\_10494+4424 & ULIRG + LINER & \citet{Imanishi07} & 10:52:23.50 & 44:08:48.00 & 0.092 \\
IRAS\_11130-2659 & ULIRG + LINER & \citet{Imanishi07} & 11:15:31.60 & -27:16:23.00 & 0.136 \\
IRAS\_11387+4116 & ULIRG + HII & \citet{Imanishi07} & 11:41:22.00 & 40:59:51.00 & 0.149 \\
IRAS\_11506+1331 & ULIRG + HII & \citet{Imanishi07} & 11:53:14.20 & 13:14:28.00 & 0.127 \\
IRAS\_12112+0305 & ULIRG + LINER & \citet{Imanishi07} & 12:13:46.00 & 02:48:38.00 & 0.073 \\
IRAS\_12359-0725 & ULIRG + LINER & \citet{Imanishi07} & 12:38:31.60 & -07:42:25.00 & 0.138 \\
IRAS\_13335-2612 & ULIRG + LINER & \citet{Imanishi07} & 13:36:22.30 & -26:27:34.00 & 0.125 \\
IRAS\_13469+5833 & ULIRG + HII & \citet{Imanishi09} & 13:48:40.20 & 58:18:52.00 & 0.158 \\
IRAS\_13509+0442 & ULIRG + HII & \citet{Imanishi07} & 13:53:31.60 & 04:28:05.00 & 0.136 \\
IRAS\_13539+2920 & ULIRG + HII & \citet{Imanishi07} & 13:56:10.00 & 29:05:35.00 & 0.108 \\
IRAS\_14060+2919 & ULIRG + HII & \citet{Imanishi07} & 14:08:19.00 & 29:04:47.00 & 0.117 \\
IRAS\_14252-1550 & ULIRG + LINER & \citet{Imanishi07} & 14:28:01.10 & -16:03:39.00 & 0.149 \\
IRAS\_14348-1447 & Sy1 & \citet{Imanishi07} & 14:37:38.30 & -15:00:23.00 & 0.083 \\
IRAS\_15206+3342 & Sy2 & \citet{Imanishi07} & 15:22:38.00 & 33:31:36.00 & 0.125 \\
IRAS\_15225+2350 & ULIRG + HII & \citet{Imanishi07} & 15:24:43.90 & 23:40:10.00 & 0.139 \\
IRAS\_16090-0139 & ULIRG + LINER & \citet{Imanishi07} & 16:11:40.50 & -01:47:06.00 & 0.134 \\
IRAS\_16300+1558 & Sy2 & \citet{Imanishi09} & 16:32:21.40 & 15:51:46.00 & 0.242 \\
IRAS\_16333+4630 & ULIRG + LINER & \citet{Imanishi09} & 16:34:52.60 & 46:24:53.00 & 0.191 \\
IRAS\_16474+3430 & ULIRG + HII & \citet{Imanishi07} & 16:49:14.20 & 34:25:10.00 & 0.111 \\
IRAS\_16487+5447 & ULIRG + LINER & \citet{Imanishi07} & 16:49:47.00 & 54:42:35.00 & 0.104 \\
IRAS\_17028+5817 & ULIRG + LINER & \citet{Imanishi07} & 17:03:41.90 & 58:13:45.00 & 0.106 \\
IRAS\_17068+4027 & ULIRG + HII & \citet{Imanishi09} & 17:08:31.90 & 40:23:28.00 & 0.179 \\
IRAS\_20414-1651 & ULIRG + HII & \citet{Imanishi07} & 20:44:18.20 & -16:40:16.00 & 0.086 \\
IRAS\_21208-0519 & ULIRG + HII & \citet{Imanishi07} & 21:23:29.10 & -05:06:56.00 & 0.130 \\
IRAS\_21329-2346 & ULIRG + LINER & \citet{Imanishi07} & 21:35:45.80 & -23:32:35.00 & 0.125 \\
IRAS\_22206-2715 & ULIRG + HII & \citet{Imanishi07} & 22:23:28.90 & -27:00:04.00 & 0.132 \\
IRAS\_22491-1808 & ULIRG + HII & \citet{Imanishi07} & 22:51:49.20 & -17:52:23.00 & 0.076 \\
IRAS\_23129+2548 & ULIRG + LINER & \citet{Imanishi09} & 23:15:21.40 & 26:04:32.00 & 0.179 \\
IRAS\_23234+0946 & ULIRG + LINER & \citet{Imanishi07} & 23:25:56.20 & 10:02:49.00 & 0.128 \\
LH\_H901A & QSO2 & \citet{Sturm06} & 10:52:52.80 & 57:29:00.00 & 0.205 \\
M-2-40-4 & Sy1.9 & \citet{Wu09} & 15:48:24.90 & -13:45:28.00 & 0.025 \\
M-5-13-17 & Sy1.5 & \citet{Wu09} & 05:19:35.80 & -32:39:28.00 & 0.012 \\
M+0-29-23 & Sy2 & \citet{Wu09} & 11:21:12.20 & -02:59:03.00 & 0.025 \\
MIPS180$^a$ & ULIRG & \citet{yan07} & 17:15:43.54 & 58:35:31.20 & 2.470 \\
MIPS562 & IRgal & \citet{Dasyra09} & 17:12:39.60 & 58:59:55.10 & 0.540 \\
MIPS8040 & IRgal & \citet{Dasyra09} & 17:13:12.00 & 60:08:40.20 & 0.759 \\
MIPS8242$^a$ & ULIRG & \citet{yan07} & 17:14:33.17 & 59:39:11.20 & 2.450 \\
MIPS15755 & IRgal & \citet{Dasyra09} & 17:18:34.90 & 59:45:34.10 & 0.736 \\
MIPS22307 & IRgal & \citet{Dasyra09} & 17:19:51.40 & 58:42:22.80 & 0.700 \\
MIPS22352 & IRgal & \citet{Dasyra09} & 17:21:47.70 & 58:53:55.90 & 0.656 \\
Mrk52 & SB & \citet{Brandl06} & 12:25:42.67 & 00:34:20.40 & 0.007 \\
Mrk273 & Sy2 & \citet{Wu09} & 13:44:42.10 & 55:53:13.00 & 0.038 \\
Mrk334 & Sy1.8 & \citet{Deo07} & 00:03:09.60 & 21:57:37.00 & 0.022 \\
Mrk471 & Sy1.8 & \citet{Deo07} & 14:22:55.40 & 32:51:03.00 & 0.034 \\
Mrk609 & Sy1.8 & \citet{Deo07} & 03:25:25.30 & -06:08:38.00 & 0.034 \\
Mrk622 & Sy2 & \citet{Deo07} & 08:07:41.00 & 39:00:15.00 & 0.023 \\
Mrk883 & Sy1.9 & \citet{Deo07} & 16:29:52.90 & 24:26:38.00 & 0.038 \\
Mrk938 & Sy2 & \citet{Wu09} & 00:11:06.50 & -12:06:26.00 & 0.020 \\
Mrk1066 & Sy2 & \citet{Shi06} & 02:59:58.60 & 36:49:14.00 & 0.012 \\
Murphy3 & SMG & \citet{Murphy09} & 12:36:03.25 & 62:11:10.80 & 0.638 \\
Murphy8 & SMG & \citet{Murphy09} & 12:36:22.48 & 62:15:44.30 & 0.639 \\
Murphy22 & SMG & \citet{Murphy09} & 12:37:34.52 & 62:17:23.20 & 0.641 \\
NGC513 & Sy2 & \citet{Wu09} & 01:24:26.80 & 33:47:58.0 & 0.02 \\
NGC520 & SB + Sy1.8 & \citet{Brandl06} & 01:24:35.07 & 03:47:32.7 & 0.0071 \\
NGC660 & SB & \citet{Brandl06} & 01:43:02.45 & 13:38:44.4 & 0.0029 \\
NGC1056 & Sy2 & \citet{Wu09} & 02:42:48.30 & 28:34:27.00 & 0.005 \\
NGC1097 & SB + Sy1 & \citet{Wu09} & 02:46:19.08 & -30:16:28.00 & 0.004 \\
NGC1125 & Sy2 & \citet{Wu09} & 02:51:40.30 & -16:39:04.00 & 0.011 \\
NGC1143 & Sy2 & \citet{Wu09} & 02:55:12.20 & -00:11:01.00 & 0.029 \\
NGC1222 & SB & \citet{Brandl06} & 03:08:56.74 & -02:57:18.50 & 0.008 \\
NGC1365 & SB + Sy1.8 & \citet{Brandl06} & 03:33:36.37 & -36:08:25.50 & 0.005 \\
NGC1566 & Sy1.5 & \citet{Wu09} & 04:20:00.40 & -54:56:16.00 & 0.005 \\
NGC1614 & SB + HII & \citet{Brandl06} & 04:33:59.85 & -08:34:44.00 & 0.015 \\
NGC1667 & Sy2 & \citet{Wu09} & 04:48:37.10 & -06:19:12.00 & 0.015 \\
NGC2146 & SB & \citet{Brandl06} & 06:18:37.71 & 78:21:25.30 & 0.004 \\
NGC2273 & Sy1 & \citet{Shi06} & 06:50:08.60 & 60:50:45.00 & 0.006 \\
NGC2623 & SB & \citet{Brandl06} & 08:38:24.08 & 25:45:16.90 & 0.018 \\
NGC2992 & Sy1.9 & \citet{Wu09} & 09:45:42.00 & -14:19:35.00 & 0.008 \\
NGC3079 & Sy2 & \citet{Weedman05} & 10:01:57.80 & 55:40:47.00 & 0.004 \\
NGC3227 & Sy1.5 & \citet{Wu09} & 10:23:30.60 & 19:51:54.00 & 0.004 \\
NGC3256 & SB & \citet{Brandl06} & 10:27:51.27 & -43:54:13.80 & 0.008 \\
NGC3310 & SB + HII & \citet{Brandl06} & 10:38:45.96 & 53:30:12.00 & 0.005 \\
NGC3511 & Sy1 & \citet{Wu09} & 11:03:23.80 & -23:05:12.00 & 0.004 \\
NGC3556 & SB & \citet{Brandl06} & 11:11:30.97 & 55:40:26.80 & 0.003 \\
NGC3628 & SB + LINER & \citet{Brandl06} & 11:20:17.02 & 13:35:22.20 & 0.002 \\
NGC3786 & Sy1.8 & \citet{Deo07} & 11:39:42.50 & 31:54:33.00 & 0.009 \\
NGC3982 & Sy1.9 & \citet{Wu09} & 11:56:28.10 & 55:07:31.00 & 0.004 \\
NGC4088 & SB & \citet{Brandl06} & 12:05:34.19 & 50:32:20.50 & 0.003 \\
NGC4194 & SB + HII & \citet{Brandl06} & 12:14:09.64 & 54:31:34.60 & 0.009 \\
NGC4388 & Sy2 & \citet{Wu09} & 12:25:46.70 & 12:39:44.00 & 0.008 \\
NGC4676 & SB & \citet{Brandl06} & 12:46:10.10 & 30:43:55.00 & 0.022 \\
NGC4818 & SB & \citet{Brandl06} & 12:56:48.90 & -08:31:31.10 & 0.002 \\
NGC4945 & SB + Sy2 & \citet{Brandl06} & 13:05:27.48 & -49:28:05.60 & 0.001 \\
NGC5005 & Sy2 & \citet{Wu09} & 13:10:56.20 & 37:03:33.00 & 0.003 \\
NGC5033 & Sy1.8 & \citet{Wu09} & 13:13:27.50 & 36:35:38.00 & 0.003 \\
NGC5135 & Sy2 & \citet{Wu09} & 13:25:44.00 & -29:50:01.00 & 0.014 \\
NGC5194 & Sy2 & \citet{Wu09} & 13:29:52.70 & 47:11:43.00 & 0.002 \\
NGC5256 & Sy2 & \citet{Wu09} & 13:38:17.50 & 48:16:37.00 & 0.028 \\
NGC5674 & Sy1.9 & \citet{Shi06} & 14:33:52.20 & 05:27:30.00 & 0.025 \\
NGC5953 & Sy2 & \citet{Buchanot06} & 15:34:32.40 & 15:11:38.00 & 0.007 \\
NGC6810 & Sy2 & \citet{Wu09} & 19:43:34.40 & -58:39:21.00 & 0.007 \\
NGC6890 & Sy1.9 & \citet{Wu09} & 20:18:18.10 & -44:48:25.00 & 0.008 \\
NGC7130 & Sy1.9 & \citet{Buchanot06} & 21:48:19.50 & -34:57:05.00 & 0.016 \\
NGC7252 & SB & \citet{Brandl06} & 22:20:44.77 & -24:40:41.80 & 0.016 \\
NGC7469 & Sy1.5 & \citet{Wu09} & 23:03:15.60 & 08:52:26.00 & 0.016 \\
NGC7496 & Sy2 & \citet{Buchanot06} & 23:09:47.30 & -43:25:41.00 & 0.006 \\
NGC7582 & Sy2 & \citet{Wu09} & 23:18:23.50 & -42:22:14.00 & 0.005 \\
NGC7590 & Sy2 & \citet{Wu09} & 23:18:55.00 & -42:14:17.00 & 0.005 \\
NGC7603 & Sy1.5 & \citet{Wu09} & 23:18:56.60 & 00:14:38.00 & 0.030 \\
NGC7714 & SB + HII & \citet{Brandl06} & 23:36:14.10 & 02:09:18.60 & 0.009 \\
SDSS\_J005621.72+003235.8 & Sy2 & \citet{Zakamska08} & 00:56:21.72 & 00:32:35.80 & 0.484 \\
SJ103837.03+582214.8$^a$ & SB & \citet{Weedman06b} & 10:38:37.03 & 58:22:14.80 & 1.680 \\
SJ104217.17+575459.2$^a$ & SB & \citet{Weedman06b} & 10:42:17.17 & 57:54:59.20 & 1.910 \\
SJ104731.08+581016.1$^a$ & SB & \citet{Weedman06b} & 10:47:31.08 & 58:10:16.10 & 1.810 \\
SST172458.3+591545 & Sy1 & \citet{Hiner09} & 17:24:58.30 & 59:15:45 & 0.494 \\
SWIRE4\_J103637.18+584217.0$^a$ & ULIRG & \citet{Farrah09} & 10:36:37.18 & 58:42:17.00 & 0.970 \\
SWIRE4\_J104057.84+565238.9$^a$ & ULIRG & \citet{Farrah09} & 10:40:57.84 & 56:52:38.90 & 0.930 \\
SWIRE4\_J104117.93+595822.9$^a$ & ULIRG & \citet{Farrah09} & 10:41:17.93 & 59:58:22.90 & 0.650 \\
SWIRE4\_J104830.58+591810.2 & ULIRG & \citet{Farrah09} & 10:48:30.58 & 59:18:10.20 & 0.940 \\
SWIRE4\_J105943.83+572524.9$^a$ & ULIRG & \citet{Farrah09} & 10:59:43.83 & 57:25:24.90 & 0.800 \\
UGC5101 & Sy1 & \citet{Wu09} & 09:35:51.60 & 61:21:11.00 & 0.039 \\
UGC7064 & Sy1.9 & \citet{Wu09} & 12:04:43.30 & 31:10:38.00 & 0.025 \\
UGC12138 & Sy1.8 & \citet{Deo07} & 22:40:17.00 & 08:03:14.00 & 0.025 
\end{xtabular}
\end{center}

%% file: table_62curvefit.tex
\topcaption{Best-fit results for the 6.2~$\mu$m band utilizing \textit{curve\_fit} and equation \ref{eq:gauss} (Section \ref{sub-analysis}). A is the amplitude, $\lambda_c$ is the central wavelength and FWHM is the full width at half maximum.}
\label{tab:cfit}

\tablefirsthead{\toprule 
Source & Class & A  & Err A & $\lambda_c$ & Err $\lambda_c$ & FWHM & Err FWHM  & RMS \\
&  & (Jy/sr) & (Jy/sr) & ($\mu$m) & ($\mu$m) & &  & (\%) \\   
\midrule}

\tablehead{%
\multicolumn{9}{c}%
{{\bfseries  Continued from previous page}} \\
\toprule 
Source & Class & A  & Err A & $\lambda_c$ & Err $\lambda_c$ & FWHM & Err FWHM  & RMS \\
&  & (Jy/sr) & (Jy/sr) & ($\mu$m) & ($\mu$m) & &  & (\%) \\   
\midrule}

\tabletail{%
\midrule \multicolumn{9}{r}{{Continued from previous page}} \\ \midrule}
\tablelasttail{%
\\\midrule
\multicolumn{9}{l}{{* Objects with profile's red tail disconsidered from the fit.}} }

\begin{center}
\begin{xtabular}{lcccccccc}
3C293 & B & 0.695 & 0.049 & 6.234 & 0.006 & 0.175 & 0.015 & 7.155 \\
3C31 & A & 1.611 & 0.043 & 6.230 & 0.002 & 0.193 & 0.007 & 3.120 \\
AGN15 & B & 0.453 & 0.037 & 6.258 & 0.008 & 0.209 & 0.021 & 5.692 \\
Arp220 & A & 35.186 & 1.446 & 6.223 & 0.003 & 0.164 & 0.009 & 4.986 \\
CGCG381-051 & A & 7.201 & 0.839 & 6.227 & 0.009 & 0.149 & 0.021 & 12.127 \\
E12-G21 & A & 14.414 & 0.794 & 6.226 & 0.004 & 0.162 & 0.011 & 6.058 \\
EIRS-14 & A & 0.208 & 0.008 & 6.225 & 0.003 & 0.202 & 0.010 & 2.937 \\
EIRS-2* & A & 0.181 & 0.011 & 6.229 & 0.004 & 0.135 & 0.011 & 6.905 \\
EIRS-41 & B & 0.118 & 0.023 & 6.263 & 0.013 & 0.155 & 0.036 & 11.921 \\
GN26 & B & 0.155 & 0.012 & 6.266 & 0.006 & 0.154 & 0.014 & 5.712 \\
IC342* & A & 39.660 & 1.254 & 6.221 & 0.002 & 0.134 & 0.005 & 3.527 \\
IRAS02021-2103* & B & 3.261 & 0.165 & 6.239 & 0.005 & 0.189 & 0.011 & 4.195 \\
IRAS02480-3745* & A & 2.701 & 0.184 & 6.214 & 0.007 & 0.203 & 0.017 & 5.412 \\
IRAS03209-0806* & A & 2.506 & 0.098 & 6.209 & 0.003 & 0.147 & 0.007 & 4.738 \\
IRAS04074-2801* & A & 1.447 & 0.113 & 6.217 & 0.005 & 0.144 & 0.014 & 7.855 \\
IRAS05020-2941* & A & 1.632 & 0.200 & 6.214 & 0.007 & 0.115 & 0.017 & 13.094 \\
IRAS08591+5248* & A & 2.224 & 0.098 & 6.210 & 0.003 & 0.144 & 0.008 & 4.428 \\
IRAS10594+3818* & A & 4.092 & 0.161 & 6.203 & 0.003 & 0.142 & 0.007 & 3.221 \\
IRAS12447+3721 & A & 1.968 & 0.120 & 6.219 & 0.005 & 0.167 & 0.012 & 4.497 \\
IRAS13106-0922 & A & 1.021 & 0.127 & 6.224 & 0.010 & 0.166 & 0.025 & 15.227 \\
IRAS14121-0126* & A & 2.751 & 0.084 & 6.198 & 0.002 & 0.109 & 0.004 & 2.837 \\
IRAS14197+0813 & B & 1.203 & 0.153 & 6.233 & 0.008 & 0.121 & 0.016 & 9.285 \\
IRAS14202+2615 & A & 5.443 & 0.188 & 6.210 & 0.002 & 0.151 & 0.006 & 4.393 \\
IRAS14485-2434* & A & 2.063 & 0.130 & 6.201 & 0.004 & 0.123 & 0.010 & 5.812 \\
IRAS15043+5754* & A & 2.390 & 0.066 & 6.205 & 0.002 & 0.134 & 0.005 & 2.762 \\
IRAS21477+0502 & A & 1.702 & 0.140 & 6.221 & 0.006 & 0.159 & 0.015 & 10.208 \\
IRAS22088-1831 & A & 1.284 & 0.102 & 6.222 & 0.007 & 0.171 & 0.017 & 10.250 \\
IRAS\_00091-0738* & A & 0.812 & 0.073 & 6.229 & 0.004 & 0.083 & 0.009 & 6.277 \\
IRAS\_00456-2904* & A & 5.286 & 0.224 & 6.229 & 0.003 & 0.132 & 0.008 & 2.837 \\
IRAS\_01199-2307 & A & 1.390 & 0.339 & 6.223 & 0.017 & 0.143 & 0.042 & 12.177 \\
IRAS\_01355-1814 & B & 1.005 & 0.232 & 6.234 & 0.022 & 0.194 & 0.053 & 12.971 \\
IRAS\_01494-1845* & A & 1.656 & 0.140 & 6.203 & 0.005 & 0.111 & 0.011 & 7.864 \\
IRAS\_02411+0353* & A & 3.782 & 0.409 & 6.230 & 0.007 & 0.124 & 0.015 & 5.021 \\
IRAS\_03250+1606 & B & 2.986 & 0.095 & 6.244 & 0.002 & 0.146 & 0.006 & 2.936 \\
IRAS\_03521+0028* & A & 1.724 & 0.086 & 6.212 & 0.003 & 0.152 & 0.010 & 4.857 \\
IRAS\_08201+2801 & A & 2.126 & 0.092 & 6.224 & 0.004 & 0.191 & 0.011 & 3.149 \\
IRAS\_09039+0503* & B & 1.037 & 0.096 & 6.231 & 0.005 & 0.119 & 0.014 & 6.377 \\
IRAS\_09116+0334 & A & 2.959 & 0.127 & 6.228 & 0.003 & 0.162 & 0.009 & 5.452 \\
IRAS\_09463+8141 & A & 0.970 & 0.092 & 6.223 & 0.007 & 0.147 & 0.017 & 4.404 \\
IRAS\_09539+0857 & A & 1.073 & 0.054 & 6.229 & 0.003 & 0.141 & 0.009 & 5.355 \\
IRAS\_10190+1322* & A & 11.248 & 0.452 & 6.225 & 0.003 & 0.162 & 0.008 & 3.899 \\
IRAS\_10485-1447 & B & 1.554 & 0.120 & 6.237 & 0.006 & 0.159 & 0.014 & 7.475 \\
IRAS\_10494+4424 & B & 5.380 & 0.333 & 6.238 & 0.005 & 0.153 & 0.011 & 5.899 \\
IRAS\_11130-2659* & B & 0.785 & 0.079 & 6.238 & 0.005 & 0.107 & 0.013 & 8.459 \\
IRAS\_11387+4116 & B & 1.466 & 0.105 & 6.234 & 0.005 & 0.135 & 0.012 & 5.900 \\
IRAS\_11506+1331* & B & 3.769 & 0.150 & 6.242 & 0.003 & 0.138 & 0.007 & 2.846 \\
IRAS\_12112+0305* & A & 5.955 & 0.461 & 6.219 & 0.004 & 0.123 & 0.012 & 3.856 \\
IRAS\_12359-0725 & B & 1.091 & 0.059 & 6.240 & 0.003 & 0.139 & 0.010 & 5.229 \\
IRAS\_13335-2612* & B & 3.467 & 0.156 & 6.239 & 0.004 & 0.177 & 0.009 & 4.648 \\
IRAS\_13469+5833 & A & 1.186 & 0.036 & 6.223 & 0.002 & 0.150 & 0.005 & 2.225 \\
IRAS\_13509+0442* & B & 2.571 & 0.118 & 6.241 & 0.003 & 0.130 & 0.008 & 4.209 \\
IRAS\_13539+2920 & B & 4.008 & 0.111 & 6.241 & 0.002 & 0.142 & 0.005 & 2.954 \\
IRAS\_14060+2919* & A & 4.536 & 0.302 & 6.229 & 0.004 & 0.138 & 0.012 & 4.740 \\
IRAS\_14252-1550* & A & 1.482 & 0.092 & 6.216 & 0.004 & 0.151 & 0.012 & 5.129 \\
IRAS\_14348-1447 & B & 6.712 & 0.419 & 6.231 & 0.005 & 0.170 & 0.014 & 5.871 \\
IRAS\_15206+3342* & B & 3.895 & 0.169 & 6.235 & 0.002 & 0.117 & 0.006 & 3.030 \\
IRAS\_15225+2350* & B & 1.409 & 0.092 & 6.245 & 0.003 & 0.107 & 0.009 & 4.177 \\
IRAS\_16090-0139 & B & 3.432 & 0.139 & 6.248 & 0.003 & 0.153 & 0.008 & 3.864 \\
IRAS\_16300+1558 & A & 1.091 & 0.062 & 6.218 & 0.004 & 0.140 & 0.010 & 4.454 \\
IRAS\_16333+4630 & A & 1.812 & 0.068 & 6.218 & 0.002 & 0.122 & 0.005 & 4.118 \\
IRAS\_16474+3430* & B & 4.680 & 0.169 & 6.245 & 0.002 & 0.143 & 0.006 & 2.260 \\
IRAS\_16487+5447* & A & 1.361 & 0.116 & 6.223 & 0.004 & 0.102 & 0.010 & 5.914 \\
IRAS\_17028+5817* & B & 2.479 & 0.074 & 6.237 & 0.002 & 0.138 & 0.005 & 2.784 \\
IRAS\_17068+4027* & A & 2.280 & 0.045 & 6.211 & 0.001 & 0.153 & 0.004 & 2.082 \\
IRAS\_20414-1651 & B & 2.495 & 0.086 & 6.233 & 0.002 & 0.133 & 0.006 & 3.963 \\
IRAS\_21208-0519* & B & 2.247 & 0.206 & 6.239 & 0.007 & 0.174 & 0.021 & 8.246 \\
IRAS\_21329-2346* & B & 1.100 & 0.077 & 6.231 & 0.004 & 0.117 & 0.010 & 5.661 \\
IRAS\_22206-2715 & B & 1.472 & 0.115 & 6.235 & 0.005 & 0.118 & 0.012 & 6.316 \\
IRAS\_22491-1808 & B & 4.206 & 0.271 & 6.240 & 0.004 & 0.127 & 0.009 & 4.841 \\
IRAS\_23129+2548 & A & 1.533 & 0.111 & 6.228 & 0.006 & 0.173 & 0.016 & 9.015 \\
IRAS\_23234+0946 & A & 1.550 & 0.087 & 6.227 & 0.004 & 0.164 & 0.012 & 6.789 \\
LH\_H901A & A & 0.119 & 0.006 & 6.230 & 0.004 & 0.159 & 0.010 & 5.777 \\
M+0-29-23 & A & 25.504 & 0.864 & 6.226 & 0.003 & 0.163 & 0.007 & 4.229 \\
M-2-40-4 & A & 11.328 & 0.820 & 6.220 & 0.006 & 0.173 & 0.016 & 8.827 \\
M-5-13-17 & B & 12.566 & 1.449 & 6.233 & 0.011 & 0.250 & 0.036 & 7.829 \\
MIPS180* & C & 0.137 & 0.006 & 6.303 & 0.006 & 0.320 & 0.016 & 10.540 \\
MIPS562 & A & 0.228 & 0.019 & 6.229 & 0.007 & 0.185 & 0.020 & 6.512 \\
MIPS8040 & B & 0.174 & 0.010 & 6.235 & 0.004 & 0.132 & 0.009 & 5.225 \\
MIPS8242* & B & 0.050 & 0.004 & 6.282 & 0.006 & 0.146 & 0.014 & 9.851 \\
MIPS15755 & B & 0.223 & 0.010 & 6.231 & 0.003 & 0.156 & 0.009 & 4.730 \\
MIPS22307* & A & 0.115 & 0.007 & 6.217 & 0.003 & 0.090 & 0.006 & 4.758 \\
MIPS22352 & A & 0.324 & 0.013 & 6.230 & 0.004 & 0.225 & 0.012 & 4.023 \\
Mrk52* & A & 12.490 & 0.481 & 6.209 & 0.003 & 0.146 & 0.007 & 3.900 \\
Mrk273* & A & 11.823 & 0.976 & 6.212 & 0.005 & 0.130 & 0.013 & 7.593 \\
Mrk334 & B & 16.966 & 0.760 & 6.244 & 0.004 & 0.156 & 0.009 & 3.191 \\
Mrk471 & B & 3.212 & 0.231 & 6.245 & 0.006 & 0.181 & 0.015 & 3.970 \\
Mrk609 & A & 10.671 & 0.479 & 6.225 & 0.004 & 0.164 & 0.009 & 4.127 \\
Mrk622 & A & 2.439 & 0.234 & 6.210 & 0.007 & 0.141 & 0.016 & 6.779 \\
Mrk883 & B & 2.445 & 0.061 & 6.231 & 0.002 & 0.174 & 0.005 & 2.178 \\
Mrk938* & A & 32.861 & 0.501 & 6.215 & 0.001 & 0.134 & 0.002 & 1.422 \\
Mrk1066 & A & 19.651 & 0.593 & 6.227 & 0.003 & 0.205 & 0.008 & 3.265 \\
Murphy3 & B & 0.183 & 0.006 & 6.232 & 0.003 & 0.188 & 0.008 & 3.300 \\
Murphy8 & B & 0.088 & 0.004 & 6.242 & 0.004 & 0.171 & 0.009 & 3.463 \\
Murphy22 & A & 0.139 & 0.005 & 6.220 & 0.003 & 0.190 & 0.009 & 3.316 \\
NGC513* & A & 11.852 & 1.162 & 6.228 & 0.006 & 0.179 & 0.024 & 6.312 \\
NGC520* & B & 65.937 & 1.346 & 6.233 & 0.002 & 0.148 & 0.004 & 2.020 \\
NGC660* & A & 139.918 & 4.928 & 6.221 & 0.003 & 0.154 & 0.007 & 4.166 \\
NGC1056* & A & 33.577 & 2.062 & 6.221 & 0.005 & 0.155 & 0.012 & 5.681 \\
NGC1097* & A & 117.594 & 5.763 & 6.214 & 0.004 & 0.163 & 0.010 & 3.869 \\
NGC1125 & A & 7.325 & 0.305 & 6.218 & 0.003 & 0.139 & 0.007 & 4.936 \\
NGC1143 & B & 22.696 & 1.473 & 6.231 & 0.005 & 0.160 & 0.013 & 7.458 \\
NGC1222* & A & 26.367 & 0.807 & 6.223 & 0.002 & 0.153 & 0.006 & 3.244 \\
NGC1365 & A & 32.675 & 1.013 & 6.216 & 0.002 & 0.169 & 0.007 & 3.541 \\
NGC1566* & A & 28.350 & 1.036 & 6.219 & 0.003 & 0.147 & 0.007 & 3.827 \\
NGC1614* & A & 62.998 & 2.944 & 6.229 & 0.003 & 0.141 & 0.008 & 2.908 \\
NGC1667 & A & 22.479 & 2.130 & 6.226 & 0.008 & 0.177 & 0.021 & 10.439 \\
NGC2146* & A & 334.740 & 10.222 & 6.224 & 0.002 & 0.151 & 0.006 & 3.538 \\
NGC2273 & A & 12.619 & 0.484 & 6.229 & 0.003 & 0.198 & 0.010 & 4.383 \\
NGC2623* & A & 18.147 & 0.518 & 6.222 & 0.002 & 0.138 & 0.005 & 3.274 \\
NGC2992 & B & 21.411 & 1.301 & 6.234 & 0.005 & 0.192 & 0.015 & 6.077 \\
NGC3079* & A & 87.002 & 2.532 & 6.219 & 0.002 & 0.161 & 0.006 & 3.278 \\
NGC3227* & A & 24.795 & 1.318 & 6.218 & 0.004 & 0.167 & 0.011 & 4.418 \\
NGC3256* & A & 120.455 & 4.367 & 6.213 & 0.003 & 0.148 & 0.007 & 3.916 \\
NGC3310* & A & 39.139 & 1.175 & 6.222 & 0.002 & 0.161 & 0.006 & 3.374 \\
NGC3511 & B & 25.836 & 1.976 & 6.238 & 0.008 & 0.224 & 0.020 & 5.848 \\
NGC3556* & A & 21.479 & 0.657 & 6.207 & 0.002 & 0.149 & 0.006 & 3.398 \\
NGC3628* & A & 75.757 & 2.304 & 6.218 & 0.002 & 0.143 & 0.005 & 2.679 \\
NGC3786* & A & 4.701 & 0.695 & 6.201 & 0.010 & 0.148 & 0.028 & 7.798 \\
NGC3982 & B & 20.085 & 1.241 & 6.235 & 0.005 & 0.179 & 0.013 & 6.602 \\
NGC4088* & A & 13.764 & 0.410 & 6.215 & 0.002 & 0.145 & 0.005 & 2.587 \\
NGC4194* & A & 83.204 & 2.671 & 6.217 & 0.002 & 0.148 & 0.006 & 3.778 \\
NGC4388 & A & 18.174 & 0.864 & 6.220 & 0.004 & 0.186 & 0.011 & 4.722 \\
NGC4676* & A & 11.294 & 0.465 & 6.219 & 0.003 & 0.141 & 0.007 & 3.152 \\
NGC4818* & A & 45.248 & 1.499 & 6.219 & 0.002 & 0.135 & 0.005 & 2.490 \\
NGC4945* & A & 111.908 & 9.806 & 6.223 & 0.005 & 0.116 & 0.012 & 7.312 \\
NGC5005 & A & 17.236 & 0.666 & 6.212 & 0.003 & 0.157 & 0.008 & 4.646 \\
NGC5033 & A & 103.131 & 3.670 & 6.226 & 0.003 & 0.182 & 0.008 & 3.797 \\
NGC5135 & A & 45.839 & 1.783 & 6.225 & 0.003 & 0.156 & 0.007 & 4.074 \\
NGC5194 & A & 135.664 & 3.886 & 6.227 & 0.002 & 0.176 & 0.006 & 3.395 \\
NGC5256* & A & 18.107 & 0.654 & 6.216 & 0.002 & 0.138 & 0.006 & 3.635 \\
NGC5674 & A & 1.939 & 0.036 & 6.219 & 0.002 & 0.256 & 0.006 & 1.332 \\
NGC5953 & B & 13.449 & 0.467 & 6.239 & 0.003 & 0.165 & 0.007 & 4.183 \\
NGC6810* & A & 56.167 & 1.677 & 6.218 & 0.002 & 0.146 & 0.005 & 2.894 \\
NGC6890* & A & 11.585 & 0.861 & 6.225 & 0.005 & 0.173 & 0.017 & 6.735 \\
NGC7130 & B & 18.447 & 0.331 & 6.244 & 0.001 & 0.171 & 0.004 & 2.149 \\
NGC7252* & A & 23.719 & 0.900 & 6.219 & 0.003 & 0.149 & 0.007 & 3.234 \\
NGC7469* & A & 58.434 & 2.659 & 6.220 & 0.003 & 0.143 & 0.008 & 4.315 \\
NGC7496 & A & 13.969 & 0.896 & 6.212 & 0.005 & 0.169 & 0.013 & 8.060 \\
NGC7582* & A & 88.536 & 3.987 & 6.221 & 0.003 & 0.137 & 0.007 & 4.654 \\
NGC7590* & A & 14.060 & 1.157 & 6.220 & 0.005 & 0.154 & 0.017 & 6.391 \\
NGC7603* & A & 8.112 & 0.852 & 6.193 & 0.007 & 0.152 & 0.021 & 9.211 \\
NGC7714* & A & 28.309 & 1.152 & 6.222 & 0.003 & 0.138 & 0.006 & 2.478 \\
SDSS\_J005621.72+003235.8* & C & 0.948 & 0.096 & 6.325 & 0.014 & 0.392 & 0.053 & 10.465 \\
SJ103837.03+582214.8 & A & 0.222 & 0.026 & 6.193 & 0.011 & 0.216 & 0.033 & 12.454 \\
SJ104217.17+575459.2 & B & 0.205 & 0.013 & 6.241 & 0.006 & 0.247 & 0.019 & 4.945 \\
SJ104731.08+581016.1 & A & 0.251 & 0.022 & 6.215 & 0.007 & 0.185 & 0.021 & 8.785 \\
SST172458.3+591545 & B & 0.115 & 0.008 & 6.233 & 0.005 & 0.160 & 0.014 & 7.053 \\
SWIRE4\_J103637.18+584217.0* & C & 0.229 & 0.014 & 6.330 & 0.008 & 0.275 & 0.020 & 9.759 \\
SWIRE4\_J104057.84+565238.9 & A & 0.099 & 0.006 & 6.220 & 0.005 & 0.192 & 0.014 & 5.687 \\
SWIRE4\_J104117.93+595822.9 & A & 0.105 & 0.003 & 6.221 & 0.003 & 0.200 & 0.008 & 2.939 \\
SWIRE4\_J104830.58+591810.2* & A & 0.208 & 0.008 & 6.213 & 0.003 & 0.164 & 0.009 & 4.055 \\
SWIRE4\_J105943.83+572524.9* & A & 0.085 & 0.009 & 6.206 & 0.007 & 0.136 & 0.017 & 10.854 \\
UGC5101* & A & 14.196 & 0.681 & 6.213 & 0.003 & 0.150 & 0.009 & 5.359 \\
UGC7064* & B & 11.935 & 1.186 & 6.245 & 0.011 & 0.227 & 0.029 & 11.408 \\
UGC12138 & A & 2.622 & 0.118 & 6.222 & 0.003 & 0.155 & 0.009 & 3.400 
\end{xtabular}
\end{center}

%% file: table_Tidx.tex
\topcaption{Values of r$_{PDR}$ and F$_{25}$/F$_{20}$ ratio for sources from the MIR\_SB sample (Spitzer/IRS ATLAS, version 1.0).}\label{tab:Tidx}

\tablefirsthead{\toprule 
ID &
\multicolumn{1}{c}{r$_{PDR}$} &
\multicolumn{1}{c}{Err r$_{PDR}$} &
\multicolumn{1}{c}{F$_{25}$/F$_{20}$} &
\multicolumn{1}{c}{Err F$_{25}$/F$_{20}$}

 \\ \midrule}

\tablehead{%
\multicolumn{5}{c}%
{{\bfseries  Continued from previous page}} \\
\toprule
ID &
\multicolumn{1}{c}{r$_{PDR}$} &
\multicolumn{1}{c}{Err r$_{PDR}$} &
\multicolumn{1}{c}{F$_{25}$/F$_{20}$} &
\multicolumn{1}{c}{Err F$_{25}$/F$_{20}$}  \\ \midrule}
\tabletail{%
\midrule \multicolumn{5}{r}{{Continued on next page}} \\ \midrule}
\tablelasttail{%
\\\midrule
\multicolumn{5}{r}{{}} }

\begin{center}
\begin{xtabular}{lcccc}

3C293	&	0.201	&	0.002	&	1.273	&	0.016	\\
3C31	&	0.277	&	0.002	&	1.017	&	0.019	\\
AGN15	&	0.168	&	0.009	&	1.521	&	0.051	\\
Arp220	&	0.268	&	0.001	&	3.89	&	0.009	\\
CGCG381-051	&	0.329	&	0.013	&	1.088	&	0.022	\\
E12-G21	&	0.328	&	0.008	&	1.162	&	0.035	\\
EIRS-14	&	0.485	&	0.038	&	---	&	---	\\
EIRS-2	&	0.556	&	0.019	&	---	&	---	\\
EIRS-41	&	0.236	&	0.044	&	0.789	&	0.149	\\
GN26	&	0.617	&	0.02	&	---	&	---	\\
IC342	&	0.413	&	0.001	&	2.243	&	0.004	\\
IRAS02021-2103	&	0.226	&	0.002	&	1.873	&	0.007	\\
IRAS02480-3745	&	0.472	&	0.005	&	3.205	&	0.029	\\
IRAS03209-0806	&	0.379	&	0.003	&	1.914	&	0.011	\\
IRAS04074-2801	&	0.319	&	0.002	&	2.854	&	0.021	\\
IRAS05020-2941	&	0.396	&	0.003	&	3.134	&	0.019	\\
IRAS08591+5248	&	0.51	&	0.003	&	1.866	&	0.018	\\
IRAS10594+3818	&	0.497	&	0.002	&	2.128	&	0.01	\\
IRAS12447+3721	&	0.355	&	0.003	&	2.348	&	0.016	\\
IRAS13106-0922	&	0.462	&	0.003	&	4.878	&	0.108	\\
IRAS14121-0126	&	0.448	&	0.003	&	2.045	&	0.013	\\
IRAS14197+0813	&	0.346	&	0.004	&	1.974	&	0.013	\\
IRAS14202+2615	&	0.339	&	0.002	&	1.99	&	0.006	\\
IRAS14485-2434	&	0.293	&	0.002	&	1.869	&	0.011	\\
IRAS15043+5754	&	0.529	&	0.002	&	2.936	&	0.024	\\
IRAS21477+0502	&	0.251	&	0.001	&	1.852	&	0.006	\\
IRAS22088-1831	&	0.263	&	0.001	&	3.281	&	0.013	\\
IRAS\_00091-0738	&	0.25	&	0.003	&	2.961	&	0.015	\\
IRAS\_00456-2904	&	0.529	&	0.003	&	2.039	&	0.012	\\
IRAS\_01199-2307	&	0.3	&	0.003	&	2.537	&	0.016	\\
IRAS\_01355-1814	&	0.197	&	0.004	&	2.834	&	0.024	\\
IRAS\_01494-1845	&	0.458	&	0.006	&	2.009	&	0.023	\\
IRAS\_02411+0353	&	0.38	&	0.003	&	1.601	&	0.007	\\
IRAS\_03250+1606	&	0.45	&	0.003	&	2.056	&	0.019	\\
IRAS\_03521+0028	&	0.354	&	0.004	&	2.7	&	0.018	\\
IRAS\_08201+2801	&	0.343	&	0.004	&	2.214	&	0.012	\\
IRAS\_09039+0503	&	0.371	&	0.006	&	2.87	&	0.036	\\
IRAS\_09116+0334	&	0.529	&	0.004	&	1.758	&	0.019	\\
IRAS\_09463+8141	&	0.38	&	0.008	&	3.262	&	0.077	\\
IRAS\_09539+0857	&	0.304	&	0.004	&	3.438	&	0.035	\\
IRAS\_10190+1322	&	0.559	&	0.001	&	1.926	&	0.009	\\
IRAS\_10485-1447	&	0.19	&	0.003	&	2.572	&	0.02	\\
IRAS\_10494+4424	&	0.523	&	0.003	&	2.551	&	0.02	\\
IRAS\_11130-2659	&	0.279	&	0.004	&	3.119	&	0.029	\\
IRAS\_11387+4116	&	0.479	&	0.006	&	1.986	&	0.023	\\
IRAS\_11506+1331	&	0.287	&	0.048	&	2.274	&	0.011	\\
IRAS\_12112+0305	&	0.403	&	0.003	&	2.637	&	0.014	\\
IRAS\_12359-0725	&	0.168	&	0.004	&	2.238	&	0.013	\\
IRAS\_13335-2612	&	0.493	&	0.003	&	2.091	&	0.02	\\
IRAS\_13469+5833	&	0.346	&	0.006	&	2.535	&	0.012	\\
IRAS\_13509+0442	&	0.458	&	0.004	&	2.021	&	0.018	\\
IRAS\_13539+2920	&	0.522	&	0.003	&	2.274	&	0.018	\\
IRAS\_14060+2919	&	0.499	&	0.003	&	1.757	&	0.011	\\
IRAS\_14252-1550	&	0.39	&	0.005	&	2.212	&	0.028	\\
IRAS\_14348-1447	&	0.346	&	0.003	&	2.754	&	0.014	\\
IRAS\_15206+3342	&	0.239	&	0.003	&	1.457	&	0.005	\\
IRAS\_15225+2350	&	0.219	&	0.041	&	2.097	&	0.01	\\
IRAS\_16090-0139	&	0.291	&	0.002	&	2.336	&	0.011	\\
IRAS\_16300+1558	&	0.269	&	0.04	&	3.062	&	0.02	\\
IRAS\_16333+4630	&	0.41	&	0.004	&	2.124	&	0.018	\\
IRAS\_16474+3430	&	0.467	&	0.003	&	2.084	&	0.013	\\
IRAS\_16487+5447	&	0.354	&	0.003	&	2.83	&	0.02	\\
IRAS\_17028+5817	&	0.513	&	0.004	&	2.55	&	0.027	\\
IRAS\_17068+4027	&	0.294	&	0.003	&	2.013	&	0.01	\\
IRAS\_20414-1651	&	0.417	&	0.006	&	3.185	&	0.025	\\
IRAS\_21208-0519	&	0.468	&	0.004	&	2.041	&	0.02	\\
IRAS\_21329-2346	&	0.395	&	0.005	&	3.13	&	0.038	\\
IRAS\_22206-2715	&	0.405	&	0.006	&	3.139	&	0.038	\\
IRAS\_22491-1808	&	0.352	&	0.029	&	2.818	&	0.014	\\
IRAS\_23129+2548	&	0.342	&	0.003	&	2.766	&	0.014	\\
IRAS\_23234+0946	&	0.289	&	0.005	&	2.38	&	0.018	\\
LH\_H901A	&	0.217	&	0.011	&	---	&	---	\\
M+0-29-23	&	0.526	&	0.009	&	1.332	&	0.023	\\
M-2-40-4	&	0.164	&	0.004	&	1.029	&	0.014	\\
M-5-13-17	&	0.237	&	0.009	&	1.054	&	0.021	\\
MIPS15755	&	0.373	&	0.026	&	---	&	---	\\
MIPS180	&	0.057	&	0.044	&	---	&	---	\\
MIPS22307	&	0.286	&	0.032	&	---	&	---	\\
MIPS22352	&	0.534	&	0.041	&	---	&	---	\\
MIPS562	&	0.553	&	0.054	&	---	&	---	\\
MIPS8040	&	0.293	&	0.031	&	---	&	---	\\
Mrk1066	&	0.291	&	0.002	&	1.489	&	0.004	\\
Mrk273	&	0.22	&	0.002	&	2.491	&	0.008	\\
Mrk334	&	0.326	&	0.002	&	1.361	&	0.004	\\
Mrk471	&	0.427	&	0.005	&	1.224	&	0.014	\\
Mrk52	&	0.318	&	0.005	&	1.142	&	0.005	\\
Mrk609	&	0.47	&	0.003	&	1.33	&	0.008	\\
Mrk622	&	0.199	&	0.004	&	1.325	&	0.006	\\
Mrk883	&	0.247	&	0.004	&	1.396	&	0.008	\\
Mrk938	&	0.498	&	0.005	&	1.834	&	0.019	\\
Murphy22	&	0.423	&	0.025	&	---	&	---	\\
Murphy3	&	0.382	&	0.025	&	---	&	---	\\
Murphy8	&	0.391	&	0.038	&	---	&	---	\\
NGC1056	&	0.662	&	0.007	&	1.394	&	0.037	\\
NGC1097	&	0.572	&	0.001	&	1.274	&	0.004	\\
NGC1125	&	0.303	&	0.061	&	1.368	&	0.022	\\
NGC1143	&	0.49	&	0.009	&	---	&	---	\\
NGC1222	&	0.436	&	0.002	&	1.461	&	0.005	\\
NGC1365	&	0.173	&	0.001	&	1.562	&	0.003	\\
NGC1566	&	0.422	&	0.003	&	1.053	&	0.005	\\
NGC1614	&	0.399	&	0.001	&	1.288	&	0.003	\\
NGC1667	&	0.626	&	0.009	&	1.304	&	0.045	\\
NGC2146	&	0.656	&	0.012	&	2.007	&	0.003	\\
NGC2273	&	0.23	&	0.001	&	1.206	&	0.004	\\
NGC2623	&	0.442	&	0.002	&	2.585	&	0.011	\\
NGC2992	&	0.21	&	0.005	&	1.053	&	0.012	\\
NGC3079	&	0.651	&	0.043	&	2.312	&	0.011	\\
NGC3227	&	0.172	&	0.002	&	1.067	&	0.004	\\
NGC3256	&	0.414	&	0.001	&	1.735	&	0.002	\\
NGC3310	&	0.557	&	0.002	&	1.62	&	0.007	\\
NGC3511	&	0.701	&	0.013	&	1.242	&	0.062	\\
NGC3556	&	0.607	&	0.003	&	1.881	&	0.014	\\
NGC3628	&	0.656	&	0.029	&	2.311	&	0.009	\\
NGC3786	&	0.288	&	0.003	&	1.241	&	0.008	\\
NGC3982	&	0.568	&	0.009	&	1.18	&	0.028	\\
NGC4088	&	0.539	&	0.004	&	1.542	&	0.012	\\
NGC4194	&	0.477	&	0.001	&	1.634	&	0.004	\\
NGC4388	&	0.196	&	0.004	&	1.301	&	0.011	\\
NGC4676	&	0.635	&	0.003	&	1.706	&	0.01	\\
NGC4818	&	0.355	&	0.001	&	1.298	&	0.003	\\
NGC4945	&	0.598	&	0.001	&	4.55	&	0.01	\\
NGC5005	&	0.457	&	0.007	&	1.514	&	0.033	\\
NGC5033	&	0.562	&	0.002	&	1.162	&	0.005	\\
NGC5135	&	0.477	&	0.004	&	1.376	&	0.012	\\
NGC513	&	0.465	&	0.014	&	1.046	&	0.037	\\
NGC5194	&	0.588	&	0.001	&	1.26	&	0.004	\\
NGC520	&	0.619	&	0.01	&	2.521	&	0.007	\\
NGC5256	&	0.54	&	0.008	&	1.602	&	0.028	\\
NGC5674	&	0.21	&	0.005	&	---	&	---	\\
NGC5953	&	0.556	&	0.008	&	1.485	&	0.019	\\
NGC660	&	0.558	&	0.026	&	1.934	&	0.004	\\
NGC6810	&	0.458	&	0.004	&	1.092	&	0.008	\\
NGC6890	&	0.285	&	0.008	&	1.041	&	0.02	\\
NGC7130	&	0.267	&	0.005	&	1.44	&	0.009	\\
NGC7252	&	0.678	&	0.002	&	1.564	&	0.009	\\
NGC7469	&	0.292	&	0.003	&	1.181	&	0.006	\\
NGC7496	&	0.288	&	0.005	&	1.499	&	0.012	\\
NGC7582	&	0.405	&	0.002	&	1.451	&	0.008	\\
NGC7590	&	0.623	&	0.011	&	1.206	&	0.055	\\
NGC7603	&	0.158	&	0.003	&	0.83	&	0.02	\\
NGC7714	&	0.359	&	0.002	&	1.361	&	0.004	\\
SDSS\_J005621.72+003235.8	&	0.174	&	0.005	&	---	&	---	\\
SJ103837.03+582214.8	&	0.373	&	0.063	&	---	&	---	\\
SJ104217.17+575459.2	&	0.574	&	0.124	&	---	&	---	\\
SJ104731.08+581016.1	&	0.21	&	0.042	&	---	&	---	\\
SST172458.3+591545	&	0.215	&	0.021	&	---	&	---	\\
SWIRE4\_J103637.18+584217.0	&	0.205	&	0.015	&	---	&	---	\\
SWIRE4\_J104117.93+595822.9	&	0.333	&	0.045	&	---	&	---	\\
SWIRE4\_J105943.83+572524.9	&	0.295	&	0.026	&	---	&	---	\\
UGC12138	&	0.154	&	0.002	&	1.119	&	0.005	\\
UGC5101	&	0.325	&	0.003	&	1.824	&	0.006	\\
UGC7064	&	0.345	&	0.012	&	0.966	&	0.027	
	
\end{xtabular}
\end{center}

%% file: tables_continua.tex

\topcaption{Results of the fitted amplitude (A) of the band of 6.2~$\mu$m for each continuum decomposition method.}\label{tab:cont-amp}

\tablefirsthead{\toprule 
ID &
\multicolumn{1}{c}{Plateau} &
\multicolumn{1}{c}{(i) A [Jy/sr]} &
\multicolumn{1}{c}{(ii) A [Jy/sr]} &
\multicolumn{1}{c}{(iii) A [Jy/sr]}\\

\midrule}

\tablehead{%
\multicolumn{5}{c}%
{{\bfseries  Continued from previous page}} \\
\toprule
ID &
\multicolumn{1}{c}{Plateau} &
\multicolumn{1}{c}{(i) A [Jy/sr]} &
\multicolumn{1}{c}{(ii) A [Jy/sr]} &
\multicolumn{1}{c}{(iii) A [Jy/sr]}\\

\midrule}
\tabletail{%
\midrule \multicolumn{5}{r}{{Continued on next page}} \\ \midrule}
\tablelasttail{%
\\\midrule
\multicolumn{5}{r}{{}} }

\begin{center}
\begin{xtabular}{lcccc}

3C31	&	yes	&	2.422 $\pm$ 0.078	&	2.422 $\pm$ 0.078	&	1.611 $\pm$ 0.043	\\
IRAS\_00456-2904	&	yes	&	8.190 $\pm$ 0.532	&	8.190 $\pm$ 0.532	&	5.286 $\pm$ 0.224	\\
IRAS\_09116+0334	&	yes	&	4.386 $\pm$ 0.215	&	4.386 $\pm$ 0.215	&	2.959 $\pm$ 0.127	\\
IRAS\_10190+1322	&	yes	&	15.824 $\pm$ 0.710	&	15.824 $\pm$ 0.710	&	11.248 $\pm$ 0.452	\\
IRAS\_16474+3430	&	yes	&	7.327 $\pm$ 0.449	&	7.327 $\pm$ 0.449	&	4.680 $\pm$ 0.169	\\
IRAS08591+5248	&	yes	&	3.620 $\pm$ 0.201	&	3.945 $\pm$ 0.213	&	2.224 $\pm$ 0.098	\\
IRAS14202+2615	&	yes	&	7.619 $\pm$ 0.272	&	7.619 $\pm$ 0.272	&	5.443 $\pm$ 0.188	\\
IRAS15043+5754	&	yes	&	4.105 $\pm$ 0.207	&	4.105 $\pm$ 0.207	&	2.390 $\pm$ 0.066	\\
Mrk1066	&	no	&	35.947 $\pm$ 0.476	&	37.757 $\pm$ 0.762	&	19.651 $\pm$ 0.593	\\
Mrk334	&	no	&	26.897 $\pm$ 2.344	&	26.897 $\pm$ 2.344	&	16.966 $\pm$ 0.760	\\
Mrk52	&	no	&	15.240 $\pm$ 0.640	&	15.240 $\pm$ 0.640	&	12.490 $\pm$ 0.481	\\
Mrk609	&	no	&	15.649 $\pm$ 0.807	&	15.649 $\pm$ 0.807	&	10.671 $\pm$ 0.479	\\
Mrk622	&	no	&	3.440 $\pm$ 0.331	&	3.945 $\pm$ 0.199	&	2.439 $\pm$ 0.479	\\
Mrk883	&	no	&	2.994 $\pm$ 0.089	&	2.994 $\pm$ 0.089	&	2.445 $\pm$ 0.061	\\
NGC1097	&	yes	&	191.185 $\pm$ 8.974	&	193.840 $\pm$ 7.407	&	117.594 $\pm$ 5.763	\\
NGC1222	&	no	&	39.608 $\pm$ 1.380	&	39.608 $\pm$ 1.380	&	26.367 $\pm$ 0.807	\\
NGC1365	&	no	&	40.451 $\pm$ 1.171	&	40.451 $\pm$ 1.171	&	32.675 $\pm$ 1.013	\\
NGC2273	&	no	&	22.736 $\pm$ 0.326	&	23.609 $\pm$ 0.496	&	12.619 $\pm$ 0.484	\\
NGC2992	&	no	&	30.599 $\pm$ 1.694	&	30.599 $\pm$ 1.694	&	21.411 $\pm$ 1.301	\\
NGC3556	&	yes	&	35.212 $\pm$ 1.338	&	35.212 $\pm$ 1.338	&	21.479 $\pm$ 0.657

\end{xtabular}
\end{center}

\clearpage

\topcaption{Results of the fitted central wavelength ($\lambda_{c}$) of the band of 6.2~$\mu$m for each continuum decomposition method.}\label{tab:cont-wl}

\tablefirsthead{\toprule 
ID &
\multicolumn{1}{c}{Plateau} &
\multicolumn{1}{c}{(i) $\lambda_{c}$ ($\mu$m)} &
\multicolumn{1}{c}{(ii) $\lambda_{c}$ ($\mu$m)} &
\multicolumn{1}{c}{(iii) $\lambda_{c}$ ($\mu$m)}\\

\midrule}

\tablehead{%
\multicolumn{5}{c}%
{{\bfseries  Continued from previous page}} \\
\toprule
ID &
\multicolumn{1}{c}{Plateau} &
\multicolumn{1}{c}{(i) $\lambda_{c}$ ($\mu$m)} &
\multicolumn{1}{c}{(ii) $\lambda_{c}$ ($\mu$m)} &
\multicolumn{1}{c}{(iii) $\lambda_{c}$ ($\mu$m)}\\

\midrule}
\tabletail{%
\midrule \multicolumn{5}{r}{{Continued on next page}} \\ \midrule}
\tablelasttail{%
\\\midrule
\multicolumn{5}{r}{{}} }

\begin{center}
\begin{xtabular}{lcccc}

3C31	&	yes	&	6.233 $\pm$ 0.003	&	6.233 $\pm$ 0.003	&	6.230 $\pm$ 0.002	\\
IRAS\_00456-2904	&	yes	&	6.238 $\pm$ 0.007	&	6.238 $\pm$ 0.007	&	6.229 $\pm$ 0.003	\\
IRAS\_09116+0334	&	yes	&	6.231 $\pm$ 0.005	&	6.231 $\pm$ 0.005	&	6.228 $\pm$ 0.003	\\
IRAS\_10190+1322	&	yes	&	6.233 $\pm$ 0.005	&	6.233 $\pm$ 0.005	&	6.225 $\pm$ 0.003	\\
IRAS\_16474+3430	&	yes	&	6.251 $\pm$ 0.006	&	6.251 $\pm$ 0.006	&	6.245 $\pm$ 0.002	\\
IRAS08591+5248	&	yes	&	6.216 $\pm$ 0.005	&	6.203 $\pm$ 0.004	&	6.210 $\pm$ 0.003	\\
IRAS14202+2615	&	yes	&	6.212 $\pm$ 0.003	&	6.212 $\pm$ 0.003	&	6.210 $\pm$ 0.002	\\
IRAS15043+5754	&	yes	&	6.214 $\pm$ 0.005	&	6.214 $\pm$ 0.005	&	6.205 $\pm$ 0.002	\\
Mrk1066	&	no	&	6.231 $\pm$ 0.001	&	6.230 $\pm$ 0.002	&	6.227 $\pm$ 0.003	\\
Mrk334	&	no	&	6.245 $\pm$ 0.009	&	6.245 $\pm$ 0.009	&	6.244 $\pm$ 0.004	\\
Mrk52	&	no	&	6.214 $\pm$ 0.003	&	6.214 $\pm$ 0.003	&	6.209 $\pm$ 0.003	\\
Mrk609	&	no	&	6.225 $\pm$ 0.005	&	6.225 $\pm$ 0.005	&	6.225 $\pm$ 0.004	\\
Mrk622	&	no	&	6.209 $\pm$ 0.008	&	6.197 $\pm$ 0.004	&	6.210 $\pm$ 0.007	\\
Mrk883	&	no	&	6.229 $\pm$ 0.003	&	6.229 $\pm$ 0.003	&	6.231 $\pm$ 0.002	\\
NGC1097	&	yes	&	6.217 $\pm$ 0.004	&	6.213 $\pm$ 0.003	&	6.214 $\pm$ 0.004	\\
NGC1222	&	no	&	6.227 $\pm$ 0.003	&	6.227 $\pm$ 0.003	&	6.223 $\pm$ 0.002	\\
NGC1365	&	no	&	6.216 $\pm$ 0.002	&	6.216 $\pm$ 0.002	&	6.216 $\pm$ 0.002	\\
NGC2273	&	no	&	6.232 $\pm$ 0.001	&	6.232 $\pm$ 0.001	&	6.229 $\pm$ 0.003	\\
NGC2992	&	no	&	6.237 $\pm$ 0.005	&	6.237 $\pm$ 0.005	&	6.234 $\pm$ 0.005	\\
NGC3556	&	yes	&	6.213 $\pm$ 0.003	&	6.213 $\pm$ 0.003	&	6.207 $\pm$ 0.002

\end{xtabular}
\end{center}

\clearpage

\topcaption{Results of the fitted FWHM of the band of 6.2~$\mu$m for each continuum decomposition method.}\label{tab:cont-fwhm}

\tablefirsthead{\toprule 
ID &
\multicolumn{1}{c}{Plateau} &
\multicolumn{1}{c}{(i) FWHM} &
\multicolumn{1}{c}{(ii) FWHM} &
\multicolumn{1}{c}{(iii) FWHM}\\

\midrule}

\tablehead{%
\multicolumn{5}{c}%
{{\bfseries  Continued from previous page}} \\
\toprule
ID &
\multicolumn{1}{c}{Plateau} &
\multicolumn{1}{c}{(i) FWHM} &
\multicolumn{1}{c}{(ii) FWHM} &
\multicolumn{1}{c}{(iii) FWHM}\\

\midrule}
\tabletail{%
\midrule \multicolumn{5}{r}{{Continued on next page}} \\ \midrule}
\tablelasttail{%
\\\midrule
\multicolumn{5}{r}{{}} }

\begin{center}
\begin{xtabular}{lcccc}
3C31	&	yes	&	0.233 $\pm$ 0.010	&	0.233 $\pm$ 0.010	&	0.197 $\pm$ 0.007	\\
IRAS\_00456-2904	&	yes	&	0.186 $\pm$ 0.016	&	0.186 $\pm$ 0.016	&	0.132 $\pm$ 0.008	\\
IRAS\_09116+0334	&	yes	&	0.202 $\pm$ 0.014	&	0.202 $\pm$ 0.014	&	0.162 $\pm$ 0.009	\\
IRAS\_10190+1322	&	yes	&	0.201 $\pm$ 0.011	&	0.201 $\pm$ 0.011	&	0.162 $\pm$ 0.008	\\
IRAS\_16474+3430	&	yes	&	0.192 $\pm$ 0.014	&	0.195 $\pm$ 0.014	&	0.143 $\pm$ 0.006	\\
IRAS08591+5248	&	yes	&	0.178 $\pm$ 0.008	&	0.181 $\pm$ 0.013	&	0.144 $\pm$ 0.008	\\
IRAS14202+2615	&	yes	&	0.196 $\pm$ 0.013	&	0.178 $\pm$ 0.008	&	0.151 $\pm$ 0.006	\\
IRAS15043+5754	&	yes	&	0.195 $\pm$ 0.014	&	0.196 $\pm$ 0.013	&	0.134 $\pm$ 0.005	\\
Mrk1066	&	no	&	0.269 $\pm$ 0.004	&	0.286 $\pm$ 0.007	&	0.205 $\pm$ 0.008	\\
Mrk334	&	no	&	0.209 $\pm$ 0.024	&	0.209 $\pm$ 0.024	&	0.156 $\pm$ 0.009	\\
Mrk52	&	no	&	0.168 $\pm$ 0.009	&	0.168 $\pm$ 0.009	&	0.146 $\pm$ 0.007	\\
Mrk609	&	no	&	0.203 $\pm$ 0.013	&	0.203 $\pm$ 0.013	&	0.164 $\pm$ 0.009	\\
Mrk622	&	no	&	0.170 $\pm$ 0.022	&	0.184 $\pm$ 0.013	&	0.141 $\pm$ 0.016	\\
Mrk883	&	no	&	0.196 $\pm$ 0.007	&	0.196 $\pm$ 0.007	&	0.174 $\pm$ 0.005	\\
NGC1097	&	yes	&	0.212 $\pm$ 0.013	&	0.207 $\pm$ 0.010	&	0.163 $\pm$ 0.010	\\
NGC1222	&	no	&	0.195 $\pm$ 0.009	&	0.195 $\pm$ 0.009	&	0.153 $\pm$ 0.006	\\
NGC1365	&	no	&	0.186 $\pm$ 0.007	&	0.186 $\pm$ 0.007	&	0.169 $\pm$ 0.007	\\
NGC2273	&	no	&	0.262 $\pm$ 0.005	&	0.275 $\pm$ 0.007	&	0.198 $\pm$ 0.010	\\
NGC2992	&	no	&	0.223 $\pm$ 0.015	&	0.223 $\pm$ 0.015	&	0.192 $\pm$ 0.015	\\
NGC3556	&	yes	&	0.198 $\pm$ 0.010	&	0.198 $\pm$ 0.010	&	0.149 $\pm$ 0.006

\end{xtabular}
\end{center}